%% file: conference_101719.tex
\documentclass[compsoc,conference,a4paper,10pt,times]{IEEEtran}
\IEEEoverridecommandlockouts
\input{tex/usepackage}

\input{src/acronyms}

\usepackage{cite}
\usepackage{amsmath,amssymb,amsfonts}
\usepackage{algorithmic}
\usepackage{graphicx}
\usepackage{textcomp}
\usepackage{xcolor}
\def\BibTeX{{\rm B\kern-.05em{\sc i\kern-.025em b}\kern-.08em
    T\kern-.1667em\lower.7ex\hbox{E}\kern-.125emX}}
\begin{document}
\bstctlcite{IEEEexample:BSTcontrol}

\title{SoK: A Taxonomy for Cybersecurity Incident Response Influence Factors}

\author{
  \IEEEauthorblockN{Thomas Biege\IEEEauthorrefmark{1}, Marius Brockhoff\IEEEauthorrefmark{2}\IEEEauthorrefmark{3}, Jonas Kaspereit\IEEEauthorrefmark{1}\IEEEauthorrefmark{3}, Fabian Ising\IEEEauthorrefmark{2}, Lea Gröber\IEEEauthorrefmark{4}, Sebastian Schinzel\IEEEauthorrefmark{1}\IEEEauthorrefmark{2}}
  \IEEEauthorblockA{\IEEEauthorrefmark{1}FH Münster University of Applied Sciences, Steinfurt, Germany \\ \{thomas.biege, j.kaspereit, schinzel\}@fh-muenster.de}
  \IEEEauthorblockA{\IEEEauthorrefmark{2}Fraunhofer SIT and National Research Center for Applied Cybersecurity ATHENE, Steinfurt, Germany \\ \{marius.brockhoff, fabian.ising\}@sit.fraunhofer.de }
  \IEEEauthorblockA{\IEEEauthorrefmark{3}Graduate School for Applied Research in North Rhine-Westphalia (Graduate School NRW), Bochum, Germany}
  \IEEEauthorblockA{\IEEEauthorrefmark{4}International Computer Science Institute, UC Berkeley, Berkeley, CA, USA \\ lgrober@icsi.berkeley.edu}
}

\IEEEoverridecommandlockouts
\IEEEpubid{\parbox{\columnwidth}{\copyright 2026, Thomas Biege. Under license to IEEE.\\DOI 10.1109/EuroSP68448.2026.00068~\hfill}
\hspace{\columnsep}\makebox[\columnwidth]{ }}

\maketitle

\IEEEpubidadjcol

\input{src/00_1_abstract}
\input{src/00_2_keywords}

\input{src/01_introduction}

\input{src/02_background}

\input{src/03_related_work}

\input{src/04_methodology}

\input{src/05_results}

\input{src/06_discussion}

\input{src/07_conclusions}

\input{src/08_acks}

\bibliographystyle{IEEEtran}
\bibliography{litreview_phase3_2_clean, references_clean}
\input{src/09_appendix}

\end{document}

%% file: tex/usepackage.tex
\usepackage{tikz}
\usetikzlibrary{mindmap}  %
\usetikzlibrary{matrix,positioning}
\usepackage{amsmath}
\usepackage{pgfplots}
\pgfplotsset{compat=1.17} %

\usepackage{svg}
\usepackage{mdframed} 
\usepackage{color}
\usepackage{colortbl}
\usepackage{multirow}
\usepackage{rotating}
\usepackage{booktabs} %
\usepackage{harveyballs} %

\usepackage{adjustbox,lipsum}
\usepackage{makecell}
\usepackage{array}
\usepackage{longtable}

\usepackage{tabularx}
\usepackage{listings}

\usepackage{textcomp}

\usepackage[acronym]{glossaries}
\glsdisablehyper

\usepackage{calc}
\usepackage{pdflscape}

\usepackage{subcaption}

\usepackage{xspace}

\usepackage{xcolor}

\usepackage{hyperref}
\usepackage[capitalise]{cleveref}
\usepackage{utfsym}
\usepackage{siunitx}
\usepackage{enumitem}
\usepackage[utf8]{inputenc}
\usepackage[T1]{fontenc}
\usepackage[english=usenglishmax]{hyphsubst}
\usepackage{csquotes}

\usepackage[shortcuts]{extdash} %

%% file: src/acronyms.tex
\newacronym{CERT}{CERT}{Computer Emergency Response Team}
\newacronym{CIR}{CIR}{Cyber(security) Incident Response}
\newacronym{CSIRT}{CSIRT}{Computer Security Incident Response Team}
\newacronym{CyberSA}{CyberSA}{Cybersecurity Situational Awareness}
\newacronym{DARPA}{DARPA}{Defense Advanced Research Projects Agency}
\newacronym{EU-GDPR}{EU-GDPR}{European Union General Data Protection Regulation}
\newacronym{IR}{IR}{Incident Response}
\newacronym{ISP}{ISP}{Internet Service Provider}
\newacronym{NIST}{NIST}{National Institute of Standards and Technology, an agency of the United States Department of Commerce}
\newacronym{NIST CSF}{NIST CSF}{Cybersecurity Framework}
\newacronym{NIST SP}{NIST SP}{NIST Special Publication: Guidelines, technical specifications, recommendations and reference materials}
\newacronym{SA}{SA}{Situational Awareness}
\newacronym{SOC}{SOC}{Security Operation Center}
\newacronym{SOP}{SOP}{Standard Operating Procedure}

%% file: src/00_1_abstract.tex
\begin{abstract}

Cybersecurity incident response has emerged as a critical area of interest for both researchers and practitioners.
The corpus of literature on cybersecurity incident response is expanding, yet a unified framework for systematically organizing the accumulated knowledge remains absent.
The aspects of incident response span multiple domains, including technology, human-computer interaction, organizational theory, and human factors.
A comprehensive, integrative perspective on these factors can enable researchers to identify underexplored areas and more effectively target their empirical and theoretical investigations.
Our study systematizes the factors that influence organizational preparedness for and response to cybersecurity incidents.
Through a systematic review of academic literature (n = 417) and non-scientific publications (n = 40), we derived the "Cybersecurity Incident Response Influencing Factor Taxonomy" (\textit{CIR-IF Taxonomy}).
Existing empirical findings were classified within this taxonomy, providing a comprehensive and up-to-date overview of knowledge from the period 1999 to mid-2024.
The taxonomy categories were systematically compared with seven established scientific frameworks and with the \textit{NIST Cyber Security Framework} elements referenced in the \textit{NIST Special Publication 800-61r3} incident response profile.
The results of this comparison show that the \textit{CIR-IF Taxonomy} delivers a richer, more rigorous, and more systematically organized view of the factors that drive and shape incident response.

\end{abstract}

%% file: src/00_2_keywords.tex
\begin{IEEEkeywords}
Cybersecurity Incident Response, Influencing Factors, Taxonomy, Systematization of Knowledge, Security Operation Center, CERT, Management, Human Factors, Context Factors
\end{IEEEkeywords}

%% file: src/01_introduction.tex
\section{Introduction}
\label{sec:introduction}

Approximately a decade after the first Computer Emergency Response Team (CERT) had been established in response to the \textit{Morris worm} incident in 1988~\cite{ormanMorrisWormFifteenyear2003}, the first scientific and practitioner-oriented publications on Computer Emergency Response Teams (CERTs) and Cybersecurity Incident Response (CIR) were published.
Early literature mainly provided guidance on how to organize a CERT and what the incident response process could look like.
Notable examples are the second edition of the "Handbook for Computer Security Incident Response Teams (CSIRTs)" by Moira West-Brown et al.~\cite{west-brown_handbook_1999} and "A common process model for incident response and computer forensics" by Felix C. Freiling and Bastian Schwittay~\cite{freiling2007common}.

\textbf{Body of literature.} Around 2014, the number of publications increased significantly; see \cref{fig:phase3_pub_per_year}.
Researchers and experts in the field were probably driven by the major cyberattacks that were carried out at this time\footnote{For example: 2010: Operation Aurora, Google, 2011: Sony Corp. PlayStation Network hack, 2013: CryptoLocker ransomware, 2013: data from 38 million Adobe customers leaked, 2017: WannaCry}.
Since then, both the frequency and complexity of cyberattacks worldwide have increased, reaching a maximum during the COVID-19 pandemic (see Statista.com\footnote{https://www.statista.com/forecasts/1485031/cyberattacks-annual-worldwide/}).
In parallel with these developments, the corresponding body of literature has undergone continuous and substantial expansion.
The interdisciplinary interest in this sociotechnical research domain~\cite {al_sabbagh_cybersecurity_2019} 
has grown, encompassing not only technological but also human-centered aspects.

\textbf{Frameworks.} Industry standards, such as NIST SP 800-61, and academic frameworks have existed for many years (see \cref{sec:related_work}).
The academic frameworks frequently build on these industry standards~\cite{pieterse_computer_2014, mooi_management_2016} or adapt established industry best practices, such as agile methods~\cite{naseer_framework_2018, smith_agile_2021} and maturity and performance models~\cite{pfleeger_improving_2017, bitzer_managing_2023}.
These contributions represent important progress toward structuring the inherent complexity of CIR.

\textbf{Research motivation.} The current body of research lacks an industry-agnostic, scientific systematization of CIR influencing factors that is constructed in a bottom-up manner from empirical data of prior work and consolidated into a taxonomy.
To lay the basis for future cybersecurity incident response research, we systematically review decades of fragmented scientific work and synthesize it into a taxonomy that organizes the full range of factors impacting incident response—turning scattered findings into an accessible, comprehensible structure.
This taxonomy is designed to give researchers from diverse disciplines a direct, low-friction entry point into the field and a solid foundation for further work. %

\textbf{Research questions.} We are driven by the general question: What are the predominant factors that influence the incident response process in organizations?
To address this question, it is first necessary to systematically classify the factors already identified in previous research.
The following research questions were defined to guide us through our study: 
\begin{itemize} [left=0pt] %
    \item[] \textbf{RQ 1}:  What categories of influencing factors have been examined in scientific papers?
    \item[] \textbf{RQ 2}: How can the factors that influence the response to cybersecurity incidents be categorized to provide a clear framework to advance research?
    \item[] \textbf{RQ 3}: Which categories are already encompassed by established cybersecurity standards (e.g., NIST SP 800-61 Rev. 3) and by existing academic frameworks?
\end{itemize}

\textbf{Contributions.} This study provides several contributions to the scientific community.
The most important result of our study is the \textit{CIR-IF Taxonomy}, see \cref{fig:cir-if-taxonomy_sunburst}.
\input{img/fig_CIF-IF-Taxonomy_Sunburst}

The \textit{CIR-IF Taxonomy} encompasses the \textsc{Govern}, \textsc{Respond}, and \textsc{Identify/Improvement} components of the incident response life-cycle model as depicted in Figure 2 of NIST SP 800-61r3.
It was developed through a three-stage systematic literature review comprising 457 publications from various domains, followed by a qualitative taxonomy development procedure.
The literature review was conducted in accordance with the PRISMA guidelines.

To apply the taxonomy development procedure described by Nickerson et al.~\cite{nickerson_method_2013}, we extended the existing formal taxonomy specification by an additional dimension (see \cref{appendix:taxonomy_devel}).
The final set of 105 analyzed academic studies is classified according to our taxonomy, thereby providing an accessible starting point for subsequent research.
As a secondary outcome of our study, we provide other researchers with a codebook for qualitative data analysis and a possible method for deriving potential relationships among influencing factors, which could be tested and verified in future research endeavors.
These scientific artifacts can support a more systematic understanding and classification of the types of influencing factors that may be identified in exploratory research.

The \textit{CIR-IF Taxonomy} was compared with seven existing academic frameworks and NIST SP 800-61r3 to demonstrate the need to address a knowledge gap in the systematization of CIR influencing factors.
Before scientists can start forming hypotheses and eventually theories, they need to understand the potential influencing factors.

\subsection{Ethical Considerations}
Our study was evaluated for potential harm to humans, animals, vulnerable groups, and society.
The systematic literature review involved collecting and analyzing data from existing publications reporting the results of scientific studies.
This content includes interview statements from user studies that were published under different ethical guidelines; this context was taken into account, especially in the citations.
The study design and content were evaluated by the Ethics Commission of 
FH Muenster.
Given its nature and data sources, the commission considered this study free of ethical concerns.

%% file: img/fig_CIF-IF-Taxonomy_Sunburst.tex
\begin{figure}
    \centering
    \includegraphics[width=\linewidth]{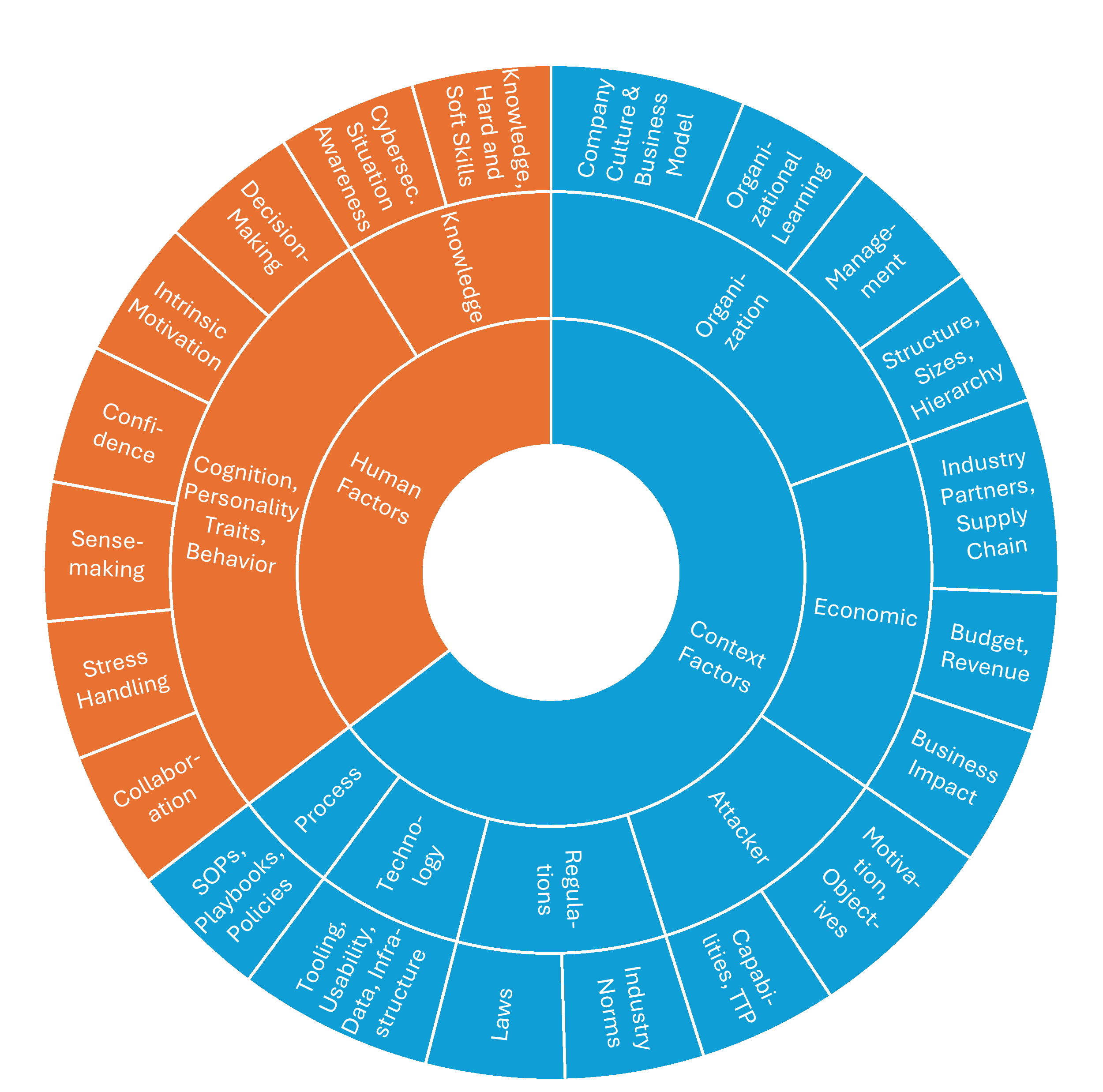}
    \caption{CIR-IF Taxonomy}
    \label{fig:cir-if-taxonomy_sunburst}
\end{figure}

%% file: src/02_background.tex
\section{Background}
\label{sec:background}

\subsection{Influencing Factor}
In this paper, an influencing factor is a variable that could positively or negatively affect an organization’s cybersecurity incident response capability.
For example, if a previous study found that the metric used to measure the performance of security analysts in a Security Operations Center leads to confusion and hinders smooth operations, we consider the performance metric an influencing factor in our analysis and classify it accordingly. %

\subsection{Cybersecurity Situational Awareness}
The Situation (or situational) Awareness model (SA) was conceptualized by Mica Endsley \cite{endsleyDesignEvaluationSituation1988}: "The perception of the elements in the environment within a volume of time and space, the comprehension of their meaning, and the projection of their status in the near future."
Endsley's model is a highly recognized framework in cognitive engineering, human factors, and command-and-control studies, categorizing SA into three hierarchical and interconnected levels: Perception, Comprehension, and Projection.
Situational Awareness (SA) presents challenges in both technical and cognitive areas.
The technical aspect involves using sensors for data collection and fusion, integrating this with external information, and delivering insights to the system operator on potential impacts. 
The cognitive aspect employs SA to aid human sensemaking in interpretation, particularly regarding cyberattacks.
In this study, this cognitive domain of \textit{Cybersecurity Situational Awareness} is used. This is reflected in our selection and deselection criteria of the systematic literature review.

\subsection{Knowledge, Skills, and Abilities}
The \textit{Knowledge, Skills, and Abilities} (KSA) framework is a foundational construct used in industrial-organizational psychology, human resources (HR), and talent management.
This framework delineates and evaluates the distinct attributes required for an individual to effectively perform the responsibilities of a specific job or role.
The KSA model is foundational to modern competency modeling, which is the practice of identifying the skills, knowledge, and behaviors required for high performance in a specific role.
\begin{itemize}
    \item \textbf{Knowledge}: The organized body of information required for a specific purpose encompasses, for instance, formal education and technical understanding.
    \item \textbf{Skills}: The learned, measurable proficiency in performing specific tasks, for instance, coding or team management.
    \item \textbf{Abilities}: The stable, underlying capacity or aptitude to perform broader functions, such as critical thinking, adaptability, or problem-solving, is also referred to as general cognitive ability.
\end{itemize}

%% file: src/03_related_work.tex
\section{Related Work}
\label{sec:related_work}

The cybersecurity profession is relatively young compared to other professions in our society and is still maturing.
In this process, standardized methods, models, and taxonomies play an important role in becoming a reliable part of an organization.
Science plays a major role in defining these standardized, reliable foundations by structuring information, as the following selected examples show.

In 2014, Pieterse~\cite{pieterse_computer_2014} developed the \textit{Computer Incident Response Framework}~(CIRF), which aims to provide a practical and holistic approach to managing a wide range of computer incidents.
This framework synthesized requirements from industry standards such as NIST, COBIT, HIPAA, and PCI DSS, and identified six core factors essential for an effective response, spanning the phases before, during, and after an incident: risk analysis, security controls, legal considerations, human aspects, digital evidence, and investigation.

Complementing this standard-driven framework, research from Mooi and Botha (2016)~\cite{mooi_management_2016} has focused on the initial establishment of response capabilities.
They developed a management model to guide the complex task of establishing a Computer Security Incident Response Team~(CSIRT).
This model utilized the \textit{IT Infrastructure Library}~(ITIL) framework, structuring requirements around five core components: people, processes, products (tools \& technologies), partners, and services.

Vielberth et al. (2020) \cite{vilberth_security_2020} conducted a systematic study to address the lack of a uniform definition for SOCs.
They proposed the adoption of an extended sociotechnical model, people, processes, technology, governance, and compliance (PPTGC), to structure the core building blocks of a state-of-the-art SOC.
To support this structure, Abd Majid and Zainol Ariffin (2021) \cite{majid_model_2021} empirically validated that human, process, and technology factors have a strong, significant relationship with the success of SOC implementation.

In response to the increasingly dynamic, sophisticated, and evolving nature of the modern cyber threat environment, recent studies have prioritized achieving \textit{agility} in incident response.
Naseer (2018) \cite{naseer_framework_2018} addressed this need by proposing a \textit{Framework of Dynamic Cybersecurity Incident Response} derived from \textit{Dynamic Capabilities Theory}.
This framework empirically investigated how organizations use real-time analytics to develop the higher-order capabilities necessary for rapid adaptation. 
Specifically, the framework identifies three crucial real-time analytics-enabled \textit{dynamic capabilities}: real-time situational awareness, dynamic risk assessment, and cyber threat intelligence generation.
The development of these capabilities enables dynamic incident response strategies, resulting in strategic and economic benefits, such as improved cost and time efficiency.

Agility played a role in Smith et al. \cite{smith_agile_2021}; the team developed the \textit{Agile Incident Response for Industrial Control Systems} (AIR4ICS) framework in 2021.
This Design Science Research (DSR) artifact integrates the Agile Scrum methodology into the incident response process, providing a modular decision-making framework intended to improve communication, information sharing, and situational awareness in a high-uncertainty context within Industrial Control Systems (ICS).

An effective incident response requires harmonizing the technological infrastructure with robust social capacities.
In 2017, Pfleeger \cite{pfleeger_improving_2017} highlighted the centrality of team dynamics and effectiveness, particularly when CSIRTs operate as \textit{Multi-Team Systems} (MTSs).
Their extensive qualitative research was incorporated into an \textit{Incident Response Performance Taxonomy} and yielded practical strategies for managers to enhance continuous learning, improve information sharing, and foster collaborative problem-solving through techniques such as cross-training.

To assist organizations in systematically assessing and developing their capabilities, \textit{Maturity Models} (MMs) are crucial managerial tools.
Addressing a gap in managerial tools applicable to organizations with low \textit{Incident Response Management} (IRM) maturity, Bitzer et al. (2022) \cite{bitzer_managing_2023} developed the \textit{Incident Response Management Maturity Model} (IRM3).
The IRM3 adopts a comprehensive sociotechnical perspective structured around four key focus areas: organization, human, tools, and processes.
The evaluation of IRM3 confirmed its utility in providing both a descriptive assessment of the status quo and a prescriptive roadmap to develop the 29 dimensions of the capabilities needed for effective IRM.

Although existing academic work provides supporting evidence, there is still no comprehensive, up-to-date framework to organize this knowledge or clearly illustrate the factors that affect various aspects of the cybersecurity incident response process and preparedness.

%% file: src/04_methodology.tex
\section{Methodology}
\label{sec:methodology}

\begin{figure}[h!]
    \centering
    \includegraphics[width=\linewidth]{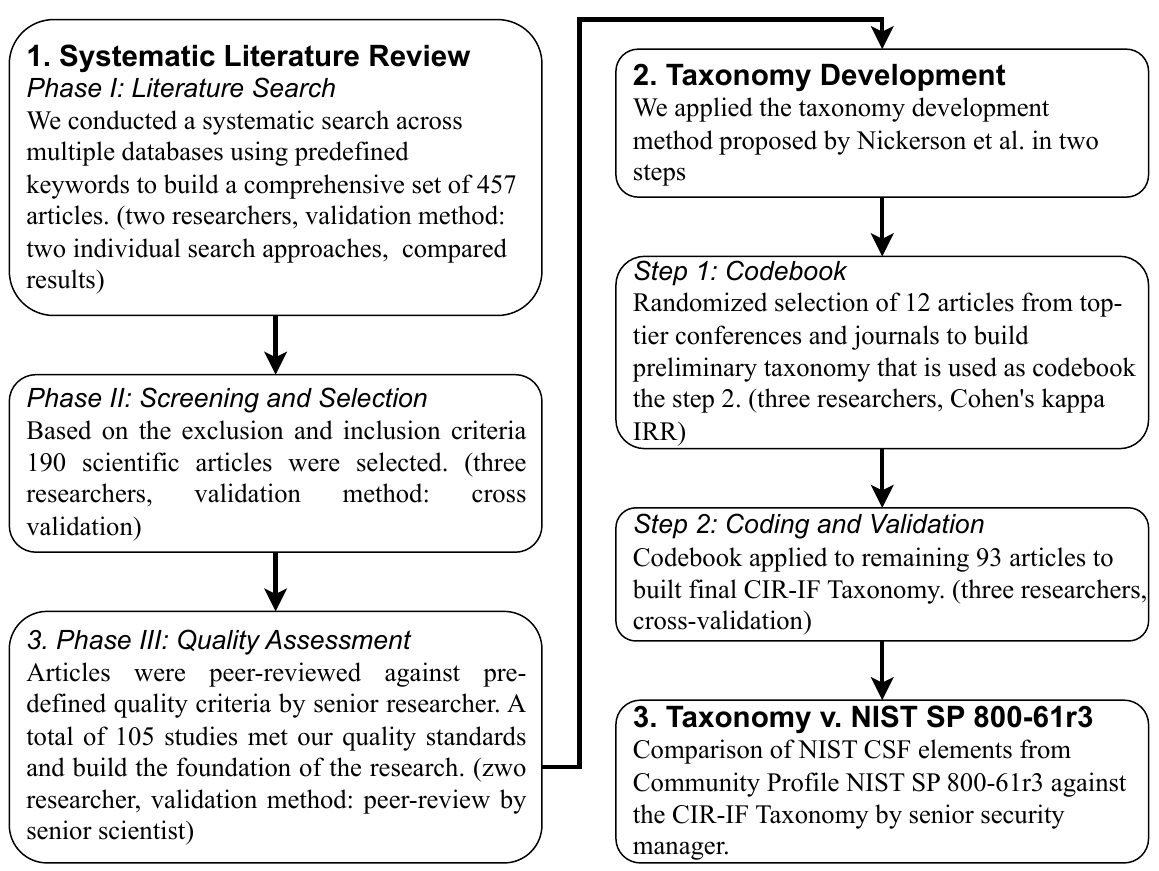}
    \caption{Research design overview.}
    \label{fig:experiment_design}
\end{figure}

The cornerstone of our work is a systematic literature review in three phases, followed by the distillation of influencing factors of the analyzed scientific work.
The influencing factors were categorized through a combination of deductive and inductive approaches, in an iterative process, to develop the cybersecurity incident response taxonomy.
\cref{fig:experiment_design} gives an overview of our research design, which we discuss in greater detail in the subsequent sections. 

\subsection{Systematic Literature Review}
To find as much literature as possible on our topic, we adopted an open approach guided by the PRISMA 2020 checklist\footnote{\url{https://www.prisma-statement.org/s/PRISMA_2020_checklist-ez8t.docx}}, starting with well-known databases and progressively filtering the results to a manageable set.
This process was carried out in three phases by a team of four researchers, referred to as A, B, C, and S (senior researcher).

\subsubsection{Phase I: Literature Search}
We designed our search strategy by systematically combining three core concepts: "incident response", "influence", and "cybersecurity".
Synonyms were generated using the \textit{Cambridge Thesaurus} online service\footnote{\url{https://dictionary.cambridge.org/us/thesaurus/}}.
In total, combining the core terms yielded 22 distinct search queries.

To systematically encompass a broad spectrum of prior work, we searched for related papers indexed by three popular databases: Google Scholar, JSTOR, and SCOPUS.
The search queries produced thousands of results, many of which were clearly out of scope based on their titles; see \cref{appendix:literature_review} for details.
To ensure rigor, two researchers independently executed the searches, retaining all papers whose titles suggested potential relevance.
This process resulted in 457 scientific and non-scientific publications.

\subsubsection{Phase II: Screening and Selection}
\label{subsec:screeningandselection}
First, we identified unique work that aligns with the definition of incident response as specified in the NIST Cybersecurity Framework \cite{national_institute_of_standards_and_technology_nist_2024}.
Consequently, we excluded studies focused solely on incident detection, analysis, or recovery, as well as unrelated domains such as kinetic incidents. 
Second, we excluded studies addressing intelligence services or military organizations.
These domains operate under different requirements and often have responsibilities, hierarchies, and structures defined by law, which could compromise the comparability and validity of our findings.
This selection and deselection was mainly based on abstract inspection, which typically sufficed to exclude articles out of scope; sometimes important parts of the article, such as the research question, methodology, and results, were read.
To mitigate reviewer bias, the process followed a rotating verification scheme: researcher A validated the results of researcher B, B verified those of C, and C reviewed the work of A.
In cases of disagreement, all three researchers discussed the case until a consensus was reached. When uncertainty persisted, the article was retained for the next evaluation phase. We selected 190 articles for further review. 

\subsubsection{Phase III: Quality Assessment}
We started with a full-text review and noticed that some papers did not meet our quality standards.
We therefore decided to conduct a formal quality assessment before developing the taxonomy.
The following criteria were applied:
(1) Peer-reviewed papers written in English and published in a recognized conference or journal, as well as doctoral theses.
(2) Studies that clearly describe their research design, including data sources, data collection, analysis methods, and results.
(3) We also excluded works whose primary contribution is the design of new processes or technologies, as such studies typically address technical innovation rather than exploring the organizational and human-specific challenges in incident response.
This assessment was conducted primarily by examining the methodology sections of the papers.
To ensure reliability, the process was carried out jointly by researcher A and S, who discussed each assessment to reach consensus.
The final set contained 105 papers.

\input{tables/tbl_litreview_phases}

\subsection{Taxonomy Development}
Our main research goal was to develop a taxonomy that provides a structured, comparable framework for organizing existing knowledge on cybersecurity incident response.
Therefore, we adapt the taxonomy development approach proposed by Nickerson et al. \cite{nickerson_method_2013} in 2013.
The formal definition of a taxonomy was enhanced to allow categories and subcategories for each dimension, see \cref{appendix:taxonomy_devel}.
In the section "Problem statement for taxonomy development", Nickerson et al. \cite{nickerson_method_2013} describe requirements for developing a taxonomy in information systems, which we adapted to qualitative research data.

The parameters for our taxonomy development process are:
\begin{itemize}
    \item \textbf{Users and use cases}: Our taxonomy should support two user groups: i.) Researchers who would like to advance research on influencing factors in cybersecurity incident response, or who will do exploratory research and need a method to classify the found objects; ii.) Practitioners in industry can use our taxonomy to understand what is needed to build up or optimize the organization's incident response capability.
    \item \textbf{Meta characteristics}: We were interested in all factors that can influence an organization's incident response capability, whether positively or negatively. Factors belong to NIST CSF domains \textsc{Govern}, \textsc{Response}, and \textsc{Identify / Improvement}.
    \item \textbf{Ending conditions}: When the \textit{Cohen's kappa} for calculating the inter-rater reliability (IRR) between two individually completed taxonomies is 0.61 or above, an acceptable level of quality is reached~\cite{landis_measurement_1977}.
\end{itemize}

For the sake of improving the \textit{flexibility} and \textit{robustness} of our taxonomy. we deviated from the \textit{concise} and \textit{comprehensive} requirements proposed by Nickerson et al. \cite{nickerson_method_2013}. 

The most significant hurdle in the development process is classifying non-discrete factors.
The method described by Nickerson et al. \cite{nickerson_method_2013} is designed for an information system, in which quantifiable parameters such as system storage size, CPU speed, or software application features can be used to define characteristics within the taxonomy.
To do the same for qualitative research results, which are often complex and ambiguous, we developed a suitable approach, as outlined in~\cref{ssec:elementsofthetaxonomy}.

\subsubsection{Step 1: Codebook}
To effectively manage an extensive volume exceeding 1000 pages of scientific content, it was imperative to commence with an intermediate taxonomy.
This taxonomy is derived from publications featured in highly respected conferences and journals, classified within the CORE ranking system as A or A*.
Six papers were randomly selected from our dataset.
As new concepts and themes emerged, we extended the taxonomy through iterative random sampling (n=6) of additional articles until no additional categories appeared and conceptual saturation was reached.
We applied a mixed deductive–inductive coding approach, beginning with a set of predefined codes (deductive) derived from existing concepts and terminology.
As we analyzed the articles, we identified recurring themes, patterns, and relationships not captured by the predefined codes.
These new insights were added iteratively, and the preliminary taxonomy was refined.

\subsubsection{Elements of the Taxonomy}
\label{ssec:elementsofthetaxonomy}
The elements, or \textit{objects} \cite{nickerson_method_2013}, collected during step 1, are the influencing factors that the corresponding research papers focused on.
They can often be found directly in the results section, sometimes as a table or graphic, or described in dedicated subsections.
If a paper has a specific topic, like "Organizational Security Learning from Incident Response" (Webb et al. \cite{webb_organizational_2017}), or "Burnout in Cybersecurity Incident Responders: Exploring the Factors that Light the Fire" (Nepal et al. \cite{nepal_burnout_2024}), it is clear that the authors' goal is to find and describe factors for burnout, respectively, for organizational learning.
Other authors used more exploratory research approaches and methods with a wider spectrum of outcomes.
A standard data collection method in this field is to gather survey results or conduct expert interviews.
We assign each identified factor to the subcategory that best represents the hypothetical reason for its influence or relationship as presented in the corresponding paper, and extract them as follows:
\begin{itemize}
    \item text segment found: "X caused by Y", or "Y leads to X", the factor is assigned to taxonomy subcategory Y
    \item text segment found: "X caused by Y caused by Z", the factor is assigned to taxonomy subcategory Z
    \item "X found", assigned to taxonomy subcategory X
\end{itemize}

We avoided interpreting the text passages that we found as much as possible.
For example, Nyre-Yu compared the needs of experts with the features of commercial SOAR solutions in her article \cite {nyre-yu_identifying_2021}, using categories such as "Situation Context".
In the evaluated data, Nyre-Yu showed the experts' need for "Situation Context" and the frequency of this feature in SOAR product marketing materials.
The researcher needs to interpret this point: the feature "Situation Context" is needed by humans and is therefore a human-related factor.
In our context, experts' needs for cybersecurity incident response were essential, and we summarized this study result as \textit{Expertise in situational context}.
This is also evident in \cref{tab:CIRIF_taxonomy_example}, which served as a codebook.

Simple but long statements were shortened without removing details of each factor and its relationship, whenever possible.
More complex statements are cited in our list of influencing factors; see \cref{tab:CIRIF_taxonomy_example}.
If the text combines several factors from different categories, we split it into individual influencing factors.

A good example is the following text segment~\cite{bulgurcu_environmental_2024}:
\begin{itemize}
    \item "Complexity of technology solutions in cybersecurity may also result in limitations to organizations [...] Nevertheless, our data unequivocally indicated that challenges pertaining to technology adaption are not primarily associated with the availability (or absence) of technology resources. Instead, these challenges are predominantly rooted in people and organization-specific factors, such as insufficient talent and training, inadequate allocation of resources, executive support, or suboptimal decision-making processes."
    \item "[...] issues related to dependencies on specific tools, over dependence on one product"
\end{itemize}
We divide statements in this manner into individual factors for each described reason/result combination.
For this specific case, the text segment led to the following five influencing factors: \\
"[...] these challenges are predominantly rooted in people and organization-specific factors, such as [...]"
\begin{itemize}
    \item[a.] "[...] insufficient talent and training, [...]"
    \item[b.] "[...] inadequate allocation of resources, [...]"
    \item[c.] "[...] executive support, [...]"
    \item[d.] "[...] or suboptimal decision-making processes."
    \item[e.] "[...] issues related to dependencies on specific tools, over dependence on one product"
\end{itemize}
This split-up of statements was a very important insight during the development process, helping us make the influencing factors more precise and less ambiguous.

\label{sec:methodology:relationsship}
Each factor was assigned to subcategories based on the underlying reason for its influence or relationship.
To better understand our dataset, we also assigned subcategories to the possible outcomes of these factors, as described or analyzed in the articles.
For example, the paper "Burnout in Cybersecurity Incident Responders: Exploring the Factors that Light the Fire" by Nepal et al. \cite{nepal_burnout_2024} explicitly looks for factors that likely cause burnout (the result).
Researchers A and B collaborated on the assignment through discussion, ensuring an objective approach by focusing strictly on the data and avoiding subjective interpretation. An example from~\cite{sundaramurthy_human_2015}:
\[ 
\overbrace{\text{\small Reduced funds allocated[...]}}^{\text{Result: Budget}}
\overbrace{\text{\small\textit{due to} management’s perception[...]}}^{\text{Reason: Management}}
\]
\[ 
\overbrace{\text{\small lack of talent[...]}}^{\text{Result: Knowledge}}
\overbrace{\text{\small \textit{due to} difficulties in affording [talented] people}}^{\text{Reason: Budget}}
\]
\\
After completing the preliminary taxonomy, we identified 136 influencing factors across the 12 top-ranked papers.
The taxonomy subcategories correspond to the coding labels in our codebook, which we used in step 2 to develop the final \textit{CIR-IF Taxonomy}. 
The process to reach a high level of inter-rater reliability is outlined in the following subsection.

\subsubsection{Reliability of the Codebook}
Due to the complexity of the taxonomy elements, several iterations, discussions, and repeated measurements of the IRR were needed to reach the qualitative attributes and requirements~\cite{nickerson_method_2013}:
\begin{itemize}
    \item \textit{mutual and collective exclusive}: An object, or influencing factor, belongs to a single subcategory. An influencing factor cannot be assigned to two subcategories in a given dimension.
    \item \textit{concise/parsimonious}: The number of dimensions and categories/characteristics should be parsimonious.
    \item \textit{robust}: To clearly and adequately differentiate the objects to create non-overlapping dimensions and characteristics. Note that this can conflict with the \textit{concise} attribute.
    \item \textit{comprehensive}: The interpretation of this attribute depends on the taxonomy development approach.
    \begin{itemize}
        \item[i.] For an empirical or inductive approach to the development of the taxonomy, the taxonomy must classify all known and considered objects (completeness).
        \item[ii.] A useful taxonomy, developed using a conceptual or deductive approach, includes all dimensions of the objects of interest.
    \end{itemize}
    \item \textit{extensible}: Inclusion of additional dimensions and new characteristics. This is an important attribute for our intention of a useful taxonomy. A static taxonomy may soon become obsolete.
    \item \textit{explanatory}: To understand objects of a specific domain, the dimensions and characteristics should provide useful explanations of the nature of the objects.
\end{itemize}

In this way, we were able to accelerate development and test the reliability of our taxonomy by having researcher S, who was not involved in the taxonomy development process, and researcher A, who was involved, perform the final categorization of the influencing factors.
The development of the preliminary taxonomy was carried out with great care and a strong emphasis on quality and reliability.
Therefore, we did three iterations in the first phase until we reached the \textit{ending condition}.
\begin{itemize}
    \item \textbf{Iteration I - Taxonomy structure}: Researcher A created the first version of the taxonomy and a list of influencing factors.

    \item \textbf{Iteration II - Improve categories and factors}: Three researchers (A, B, C) used the first version to individually assign the influencing factors. Disagreements were discussed, and inter-rater reliability (IRR) was calculated pairwise using the \textit{Cohen's kappa} per subcategory. Our $\kappa$ calculation at this stage of taxonomy development was only 0.39, a "fair" level of agreement and well below our expected quality level. Studying the differences in ratings revealed that the ambiguous naming of categories and their insufficient description led to confusion for the third researcher. Therefore, we improved the taxonomy category names, their descriptions, and, most importantly, the influencing factors before re-verifying the taxonomy's usability. Complex and ambiguous statements about relationships are cleaned up as outlined in \cref{ssec:elementsofthetaxonomy}. The \textit{Cohen's kappa} of the complete taxonomy assignment step was 0.527. This was an improvement, but does not fulfill the \textit{ending condition} of $\kappa \ge 0.61$.
    
    \item \textbf{Iteration III - Reach ending condition}: Researchers (A, B) further improved the quality to reach the \textit{ending condition}. Subcategories from earlier iterations with a low $kappa$ value were reviewed and combined if their topics were similar. Only in this way was it possible to reach the \textit{ending condition} in the development process and make subcategories mutually exclusive. For example, the assignment for the initial subcategories \textit{People Management} ($P_O = 68.75$), \textit{Human Resource Management} ($P_O = 33.33$), and \textit{Daily Work Management} ($P_O = 27.27$) had a high variation. We were forced to combine them into a single subcategory, \textit{Management} ($P_O = 77.14$). Reducing the number of subcategories by two in this case. This step helps achieve an acceptable level of quality and reliability. The new \textit{Cohen's kappa} value was 0.721, which fulfills the previously defined \textit{ending condition}.
\end{itemize}

\subsubsection{Step 2: Coding and Validation, CIR-IF Taxonomy}
The intermediate taxonomy served as a codebook for analyzing the remaining 93 papers.
During this second step, we applied the preliminary taxonomy to categorize newly identified influencing factors and refine the structure into the final \textit{CIR-IF Taxonomy}.
To validate consistency, researcher S, who was not involved in taxonomy development, coded all 105 papers in parallel with one of the original researchers.
Discrepancies between coders were discussed with all researchers and resolved collaboratively, and the taxonomy was adjusted accordingly.
Hereafter, the \textit{CIR-IF taxonomy} development process was finished.

\subsection{Comparison with NIST and existing Academic Frameworks}
The final \textit{CIR-IF Taxonomy} was compared against the well-known NIST Special Publications 800-61 Rev. 3 \cite{nelsonIncidentResponseRecommendations2025a} (a NIST CSF \textit{Community Profile}), the leading cybersecurity incident response standard at the time.
The third section of the NIST SP 800-61r3 contains two tables that comprise the so-called Community Profile.
The first table covers preparation (Govern, Identify, Protect) and lessons learned (Identify-Improvement), and the second table covers incident response (Detect, Respond, Recover).
The tables contain the identifiers of the CSF element, such as GV.OC-01, which is the first element in the subsection "Organizational Context" (OC) of the section Govern (GV).
The identifier is followed by a brief description of the CSF element, priority, and an optional recommendation or notes.
For example, the description of GV.OC-03 states "Legal, regulatory, and contractual requirements."
In the context of incident response, this CSF element was assigned to the \textit{CIR-IF Taxonomy} subsection "Laws and Regulations".

A senior security manager with two decades of experience in his profession assigned the NIST CSF elements used in the \textit{Community Profile} to the subcategories of our taxonomy.
For our evaluation, we ignored the priority of the profile's entries and included all CSF element types.
A CSF element was only assigned once to a taxonomy subcategory.

In parallel, we compared the \textit{CIR-IF Taxonomy} subcategories with the seven existing frameworks mentioned in \cref{sec:related_work}.
This process was straightforward, as the framework's elements and their meaning are well described in the scientific work.
The results are outlined in \cref{tab:taxonomy_comparison}.

%% file: tables/tbl_litreview_phases.tex
\begin{table}[htbp]
\centering
\footnotesize
\caption{Overview systematic literature review phases.}
\label{tab:lit_overview}
\begin{tabular}{@{}llc@{}}
\toprule
\textbf{Phase} & \textbf{Review}  & \textbf{Papers selected}\\
\midrule
\midrule
Phase I & Title, abstract & 457 \\
Phase II & Abstract, parts of the paper & 190 \\
Phase III & Research approach, quality assurance & 105 \\
\bottomrule
\end{tabular}
\end{table}

%% file: src/05_results.tex
\section{Results}
\label{sec:results}

\subsection{Systematic Literature Review}
We plot the number of publications by publication year of articles included in \textbf{Phase III} in \cref{fig:phase3_pub_per_year}. We see a rapid increase in publications in 2014, which continues to rise slightly until 2022\footnote{We did not include 2024 because our data collection phase ended in May 2024.}.

\input{img/fig_phase3_publications_per_year}

The most significant outcome of our systematic literature review is a comprehensive list of 105 scientific articles and dissertations that present research findings on addressing cybersecurity incidents over the past two decades.
The articles were selected in a peer-review process, the content was reviewed, and each article was assigned a human- or context-related influencing factor as well as a NIST SP 800-61r3 capability, as outlined in \cref{tab:cirif_overview_human} and \cref{tab:cirif_overview_context}.
Both tables represent a valuable reference list for self-studies.

\subsection{CIR Influencing Factor Taxonomy}
To develop the \textit{CIR-IF Taxonomy}, we advanced the original formal taxonomy description by Nickerson et al. \cite{nickerson_method_2013} to be able to present a second layer of characteristics, the subcategories of the taxonomy, see~Appendix~\ref{appendix:taxonomy_devel}.
We separated the findings into factors directly pertaining to human involvement, such as experts who respond to incidents, and factors related to the organizational context in which these experts work.
This distinction was made to make the number of factors manageable and to place the human being at the center of attention.
For example, humans are influenced by their knowledge, collaboration with other humans, and cognitive processes and abilities.
The context is defined by an organization, by environmental factors such as laws, economic pressures, or company culture, and by the motivation and capabilities of the attacker.
\input{tables/tbl_factors_overview_vertical}

\cref{tab:cirif_overview_human} and \cref{tab:cirif_overview_context} show the \textit{CIR Influencing Factor (CIR-IF) Taxonomy} (dimension, category, subcategory), the references to the corresponding articles, and the related NIST SP 800-61r3 elements.
A paper often contains influencing factors from multiple subcategories; therefore, papers are often assigned more than once.
For example, if a paper contains objects from subcategories A and B, it is assigned to A and B.

Based on the number of assignments, the top three subcategories for human factors are\footnote{percentage of human-related subcategories}:
\begin{itemize}
    \item[1.] Employees Knowledge, Soft \& Hard Skills (32\%);
    \item[2.] Cybersecurity Situation Awareness (23\%);
    \item[3.] Collaboration (19\%)
\end{itemize}
For context-related influencing factors, the top three subcategories are\footnote{percentage of context-related subcategories}:
\begin{itemize}
    \item[1.] Management (19\%);
    \item[2.] Tooling, Usability, Data Quality, IT Infra. (16\%);
    \item[3.] SOPs, Playbooks, Policies (16\%)
\end{itemize}
The numbers are mainly the result of the selection process and are influenced by the focus of the research being analyzed.

The influencing factors (used as code labels) and their corresponding textual examples are documented in our codebook. \cref{tab:if_codebook_context_1}, \cref{tab:if_codebook_context_2}, and \cref{tab:if_codebook_human}.
This codebook is suitable for use in qualitative research designs by other researchers.
For instance, it may be used in qualitative content analysis (see Mayring \cite{mayring_qualitative_2000}), the systematic coding of interview data, or the conduct of thematic analyses.

\subsubsection{Overview of Taxonomy Categories}
Unfortunately, using the \textit{CIR-IF Taxonomy} is not as straightforward as using other well-known taxonomies that use quantifiable characteristics, like the periodic table of the elements, which uses the atomic number.
It is vital to understand our intention and reasoning when choosing the subcategories.
We chose the following categories and subcategories for human-related factors:
\begin{itemize}
    \item \textbf{Knowledge}: Describes the "hard skill" knowledge about cybersecurity, cybersecurity situation awareness, tacit knowledge, level of education, or professional experience. Awareness of cybersecurity situations ranges from understanding an organization's individual risks to understanding current tactics, techniques, and procedures used by attackers. These factors do not encompass the quality or availability of books or the quality of training, but a human's ability to learn from sources of knowledge and experience.
    \item \textbf{Cognition, Personality Traits, and Behavior (CPTB)}: Covers directly observable behavior, i.e., sharing knowledge, communication with colleagues, working with others as a team, as well as aspects related to mental actions and processes that are not directly visible, like decision-making, (over-)confidence, sensemaking, but also intrinsic motivation, or personal resources and skills to cope with stress and stay mentally healthy.
\end{itemize}
 
For context-related factors, we chose the following categories and subcategories.
\begin{itemize}
    \item \textbf{Attacker}: Describes all factors belonging to the \textit{Motivation and Objectives}, or \textit{Capabilities and TTPs}, where TTPs stand for "Tactics, Techniques and Procedures"\footnote{https://attack.mitre.org/resources/faq/}.
    \item \textbf{Regulations}: Companies need to comply to \textit{Industry Norms and Standards}, \textit{Laws} as well as other \textit{Regulations} by a government. %
    \item \textbf{Economic}: The security measures and capabilities of an organization also depend on the available \textit{Budget} or on the pressure to achieve a specific \textit{Revenue} objective. The \textit{Business Impact} of an attack or of the corresponding measures plays an important role for an organization.  Additionally, the \textit{Industry Partners}, their competencies, competition, and collaboration (e.g., exchanging threat intelligence) influence incident response capabilities.
    \item \textbf{Technology}: Technology-related factors, such as \textit{Tooling}, \textit{Integration}, \textit{Data(-quality)}, \textit{Automation}, \textit{Infrastructure}, \textit{Technical Security Controls}, and \textit{Usability}.
    \item \textbf{Processes}: Processes define the work within an organization. For the cybersecurity incident response capability, \textit{Standard Operating Procedures, Playbooks, and Policies} play an important role, and are also used to implement \textit{organizational Security Controls}.
    \item \textbf{Organization}: In this category, all other factors belonging to a typical organization can be found, like \textit{Structure, Size, and Hierarchy}, \textit{Management} (i.e., people management, human resource, daily work-load management,), \textit{Organizational Learning} \cite{levitt_organizational_1988},  and \textit{Company Culture \& Business Model}, which includes unwritten rules, decision biases, cross-border cooperation, internal/external information sharing, shared business objectives, and silo thinking.
\end{itemize}

\subsubsection{A Guide to using the Taxonomy in Research}
The \textit{CIR-IF Taxonomy} is defined by the first column of \cref{tab:cirif_overview_human} and \cref{tab:cirif_overview_context} that describes the categories and subcategories of the factors.
We developed a method to structure ambiguous qualitative statements, identify the reasons mentioned in them, and classify them using a taxonomy.
The following section provides practical advice on using the codebook and taxonomy for future research.
Keep in mind that the taxonomy subcategories classify the reasons for influencing factors. %

\textbf{Decision path method.} The order of the taxonomy starts with the most specific categories, which, in our experience, are easier to decide on.
For example, suppose that a research group qualitatively analyzes interview transcripts to find more insights; they need to identify the hypothetical reason and go from top to bottom to assign the identified influencing factor to a subcategory.
Possible decision path: "Is the hypothetical reason ..."
\begin{itemize}
    \item[$\xrightarrow{\hspace*{.2em}}$] "... human-related?"
    \begin{itemize}
        \item[$\xrightarrow{\hspace*{.6em}}$] "... related to humans and their knowledge?"
        \begin{itemize}
                \item[$\xrightarrow{\hspace*{.8em}}$] "... related to humans and their knowledge with regards to formal education?"\\
        \end{itemize}
    \end{itemize}
\end{itemize}

If the right category cannot be found, it might not yet be covered in the taxonomy.
We suggest an open discussion with several researchers familiar with this field to see whether a new category has been found.

\textbf{Content-analysis method.} Another way to analyze the content of written text or interview transcripts is to use the Mayring content analysis~\cite{mayring_qualitative_2000}.
The Mayring content analysis allows for the predefinition of a coding table (deductive approach).
This coding table must be built on the \textit{CIR-IF Taxonomy} subcategories as coding labels, as shown in \cref{tab:if_codebook_human}, \cref{tab:if_codebook_context_1}, and \cref{tab:if_codebook_context_2}.
Passages on a specific factor from the corresponding papers can be used as coding examples; see~\cref{tab:if_codebook_human}, \cref{tab:if_codebook_context_1}, and \cref{tab:if_codebook_context_2}.

In this way, a researcher can assign these predefined coding labels to the text to be analyzed. If a statement in this text cannot be assigned to a coding label, it is likely that the researcher has found something unknown or missing in the taxonomy.
They should enter an open discussion about adding the new (sub-)categories to the taxonomy.

\textbf{Handling complex cases.} Sometimes it might be challenging to decide whether a factor belongs to a subcategory.
Consulting the corresponding statement from the interview transcripts or other original research data can help clarify the reasoning and context, see our approach in \cref{ssec:elementsofthetaxonomy}.
An open discussion among the researchers involved is vital to consider different viewpoints and interpretations and to understand their reasoning.
Sometimes, assigning the factor to a less specific subcategory is a pragmatic solution.
The \textit{CIR-IF Taxonomy} is a living categorization scheme that welcomes additions and adjustments.

\subsection{Comparison with Academic Frameworks}
\input{tables/tbl_taxonomy_comparison}
We evaluated the characteristics of the taxonomies and frameworks cited in \cref{sec:related_work} against the \textit{CIR-IF Taxonomy}, see \cref{tab:taxonomy_comparison}.
This evaluation includes only the characteristics that were part of the final results of these studies.
This approach is distinct from the qualitative research conducted to develop the taxonomy, in which potential influencing factors might have been identified but did not appear in the final taxonomy or framework of the cited research.
For instance, the \textit{CIR Framework} developed in \cite{pieterse_computer_2014} does not explicitly list human factors, although such factors were recognized during our coding process.

\subsection{Comparison with NIST SP 800-61r3}

\begin{figure*}[h!]
    \centering

    \begin{subfigure}[b]{0.48\linewidth}
        \centering
        \includegraphics[width=\linewidth]{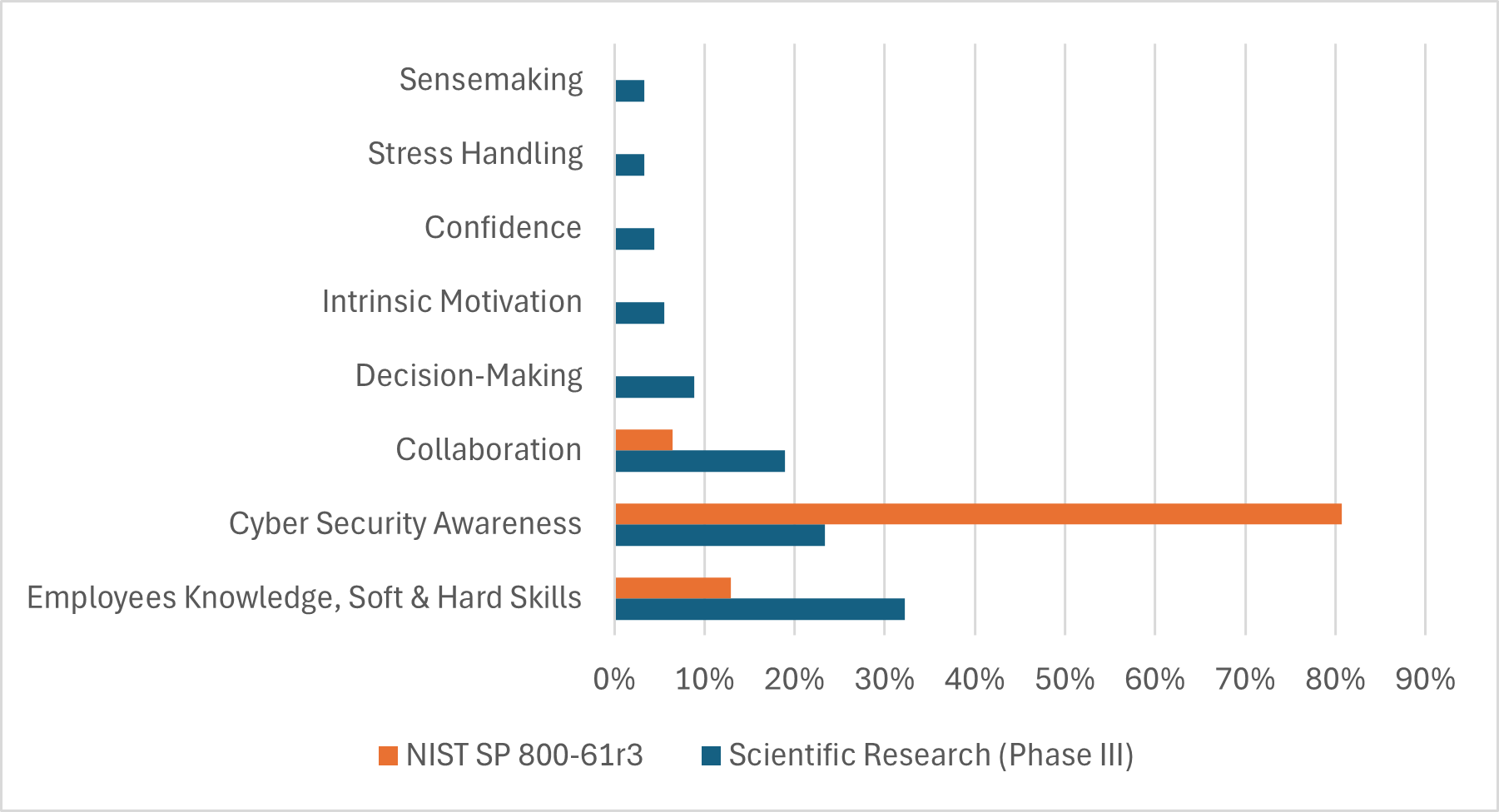}
        \caption{Human Factors: Taxonomy vs. NIST SP 800-61r3.}
        \label{fig:tax_v_nist_human}
    \end{subfigure}
    \hfill
    \begin{subfigure}[b]{0.48\linewidth}
        \centering
        \includegraphics[width=\linewidth]{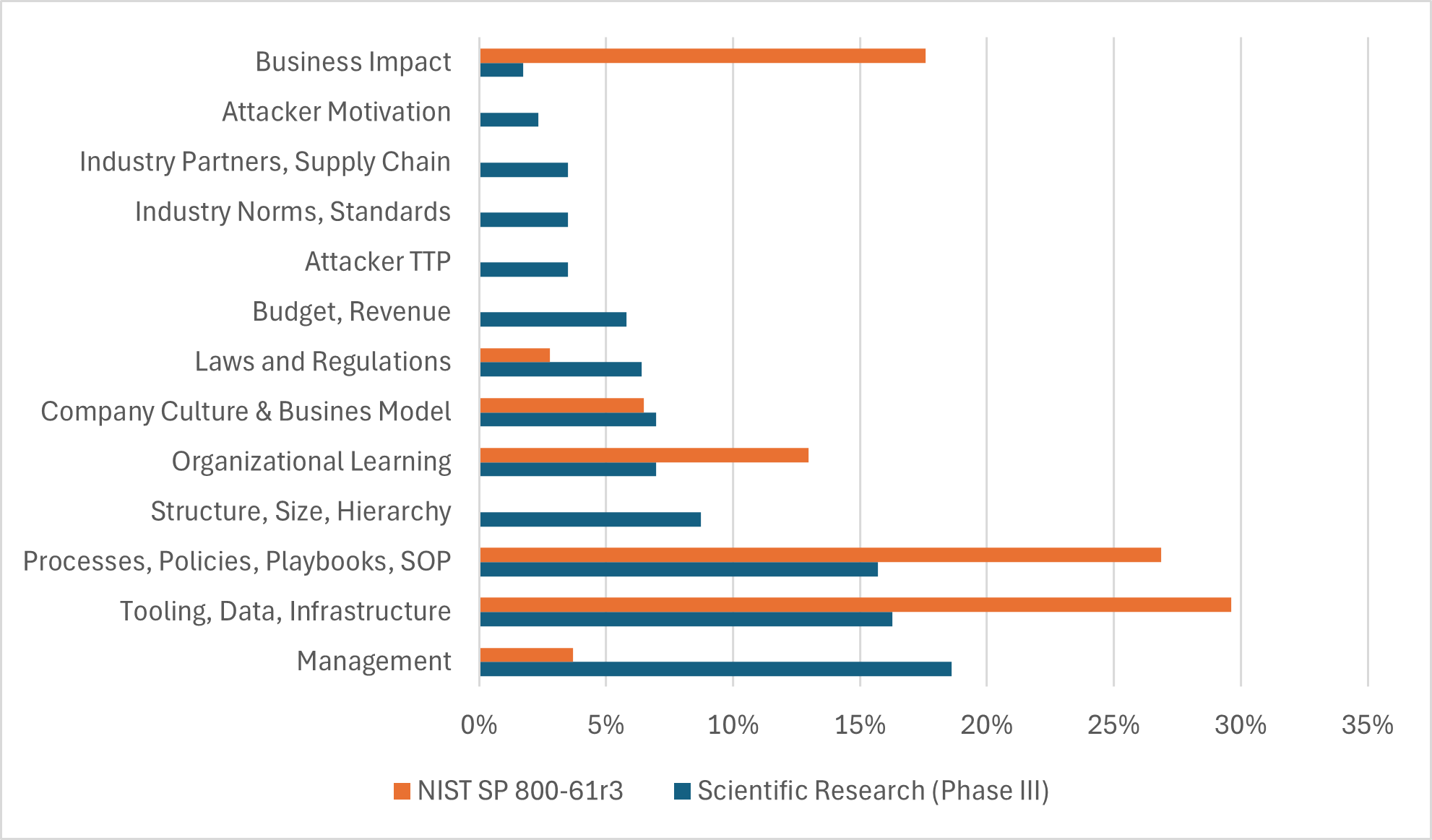}
        \caption{Context Factors: Taxonomy vs. NIST SP 800-61r3.}
        \label{fig:tax_v_nist_context}
    \end{subfigure}

    \caption{Comparison of our Taxonomy vs.\ NIST SP 800-61r3 for (a) Human Factors and (b) Context Factors.}
    \label{fig:taxonomy_nist_combined}
\end{figure*}

Mapping the elements of the NIST CSF, as outlined in NIST SP 800-61r3, to the \textit{CIR-IF Taxonomy} subcategories enabled us to pinpoint gaps and focal areas of the industry standard and assess the coverage of topics in prior scientific work.
In order to compare the thematic priorities of NIST SP 800-61r3 with those of the \textit{CIR-IF Taxonomy}, the percentage allocation for each subcategory is shown in two diagrams, \cref{fig:tax_v_nist_human}, and \cref{fig:tax_v_nist_context}.

Most of the capabilities and actions described in the NIST Special Publications document are context-related, including technology, processes, business impact, and organizational learning.
The motivation or capability of an attacker, industry partners, industry norms, the available budget, company revenue, and the company's structure, size, and hierarchy were not addressed.
Regarding human-related factors, the NIST \textit{Community Profile} covers only the top three: cybersecurity situational awareness, collaboration, and employees' knowledge and skills.
Cognitive abilities were less considered.

%% file: img/fig_phase3_publications_per_year.tex
\begin{figure}[htb]
    \centering
    \footnotesize%
    \begin{tikzpicture}
        \begin{axis}[
            ybar,
            ylabel={Publications},
            yticklabels={},
            ytick style={draw=none},
            axis x line*=left,
            axis y line*=left,
            x axis line style={white},
            y axis line style={white},
            xtick=data,
            tick style={draw=black},
            tick align=outside,
            x tick label style={rotate=90, anchor=east},
            symbolic x coords={2003, 2004, 2005, 2009, 2010, 2011, 2012, 2014, 2015, 2016, 2017, 2018, 2019, 2020, 2021, 2022, 2023},
            nodes near coords,
            width=\linewidth,
            bar width=4pt
        ]
        \addplot coordinates {
            (2003,4)
            (2004,1)
            (2005,1)
            (2009,2)
            (2010,1)
            (2011,1)
            (2012,1)
            (2014,7)
            (2015,7)
            (2016,7)
            (2017,5)
            (2018,8)
            (2019,9)
            (2020,12)
            (2021,16)
            (2022,7)
            (2023,11)
        };
        \end{axis}
    \end{tikzpicture}
    \caption{Number of Publications per Year — only years with publications from Phase III are shown (collection phase ended in May 2024)}
    \label{fig:phase3_pub_per_year}
\end{figure}

%% file: tables/tbl_factors_overview_vertical.tex
\newcolumntype{L}{>{\raggedright\arraybackslash\hspace{-.15pt}}X} %

\newcommand{\extendrow}{\rule{0pt}{8pt}}
\newcommand{\categoryrow}[1]{\rowcolor{gray!30}\multicolumn{3}{l}{\extendrow\textbf{#1}}\\[.1em]}
\newcommand{\colorrow}{\rowcolor{gray!10}}

\begin{table*}[htbp]
\centering
\footnotesize
\caption{CIR Taxonomy of Influencing Human Factors, with corresponding papers, and NIST CSF Elements from NIST SP 800-61r3.}
\label{tab:cirif_overview_human}
\begin{tabularx}{\linewidth}{p{4cm} L L}
\toprule
    \textbf{CIR-IF Taxonomy Category} &
    \textbf{Scientific Research} &
    \textbf{NIST SP 800-61r3} \\
\midrule\\[-3ex]
\categoryrow{Knowledge}
        Employees' Knowledge, Soft \& Hard Skills &
            \cite{bulgurcu_environmental_2024}, \cite{sundaramurthy_human_2015}, \cite{stevens_how_2022}, \cite{nyre-yu_identifying_2021}, \cite{kokulu_matched_2019}, \cite{al_sabbagh_cybersecurity_2019}, \cite{groenendaal_towards_2022}, \cite{line_understanding_2015}, \cite{nyre-yu_observing_2019}, \cite{pfleeger_improving_2017}, \cite{thangavelu_impact_2021}, \cite{vilberth_security_2020}, \cite{wiik_limits_2005}, \cite{pieterse_computer_2014}, \cite{majid_model_2021}, \cite{andrade_cognitive_2019}, \cite{bartnes_future_2016}, \cite{buchler_sociometrics_2018}, \cite{burkhead_phenomenological_2014}, \cite{cho_capturing_2020}, \cite{graf_decision_2016}, \cite{helkala_supporting_2018}, \cite{ho_consciousness_2021}, \cite{ioannou_cybersecurity_2019}, \cite{jaatun_framework_2009}, \cite{mases_success_2021}, \cite{oneill_cybersecurity_2021}, \cite{rollason-reese_incident_2003}, \cite{silva_measuring_2015} &
            RS.MA-02, RS.MA-03, RS.MI-01, RS.MI-02 \\
\colorrow\extendrow
Cybersecurity Situation Awareness &
            \cite{bulgurcu_environmental_2024}, \cite{sundaramurthy_tale_2014}, \cite{kokulu_matched_2019}, \cite{ahmad_how_2021}, \cite{garmon_study_2019}, \cite{mepham_dynamic_2018}, \cite{nyre-yu_observing_2019}, \cite{onwubiko_cyberops_2020}, \cite{schlette_cti-soc2m2_2021}, \cite{thangavelu_impact_2021}, \cite{pieterse_computer_2014}, \cite{andreassen_check_2023}, \cite{antunes_hybrid_2021}, \cite{ask_gamification_2021}, \cite{bartnes_line_current_2016}, \cite{gutzwiller_gaps_2020}, \cite{happa_assessing_2021}, \cite{llopis_sanchez_decision_2023}, \cite{ofte_understanding_2023}, \cite{smith_agile_2021}, \cite{thangavelu_comprehensive_2020} &
            GV.RM-05, ID.AM-*, ID.RA-01, ID.RA-02, ID.RA-03, ID.RA-04, ID.RA-05, PR.PS-04, PR.AT-01, PR.AT-02, DE.CM-*, RS.AN-03, RS.CO-02, RS.CO-03 \\
\categoryrow{Cognition, Personality Traits, and Behavior (CPTB)}
\extendrow
        Intrinsic Motivation &
        \cite{sundaramurthy_tale_2014}, \cite{kokulu_matched_2019}, \cite{rooney_what_2018}, \cite{thangavelu_impact_2021}, \cite{pieterse_computer_2014} &
        none \\
\colorrow\extendrow
        Confidence &
        \cite{nepal_burnout_2024}, \cite{bulgurcu_environmental_2024}, \cite{rooney_what_2018}, \cite{thangavelu_impact_2021} &
        none \\
\extendrow
        Stress Handling &
        \cite{nepal_burnout_2024}, \cite{sundaramurthy_human_2015}, \cite{pieterse_computer_2014} &
        none \\
\colorrow\extendrow
        Sensemaking &
         \cite{huis_human_2017}, \cite{lakshmi_sensemaking_2021}, \cite{rooney_what_2018} &
         none \\
\extendrow
        Decision-Making &
        \cite{groenendaal_towards_2022}, \cite{krichene_collective_2004}, \cite{pfleeger_improving_2017}, \cite{rooney_what_2018}, \cite{van_der_kleij_developing_2022}, \cite{pieterse_computer_2014}, \cite{helkala_supporting_2018}, \cite{smith_agile_2021} &
        none \\
\colorrow\extendrow
        Collaboration &
        \cite{sundaramurthy_turning_2016}, \cite{amador_enhancing_2020}, \cite{buchler_cyber_2018}, \cite{catota_cybersecurity_2018}, \cite{garmon_study_2019}, \cite{mepham_dynamic_2018}, \cite{moore_role_2019}, \cite{nyre-yu_observing_2019}, \cite{padayachee_coordinated_2020}, \cite{pfleeger_improving_2017}, \cite{rooney_what_2018}, \cite{vilberth_security_2020}, \cite{pieterse_computer_2014}, \cite{majid_model_2021}, \cite{cho_capturing_2020}, \cite{rollason-reese_incident_2003}, \cite{smith_agile_2021} &
        RS.CO-02, RS.CO-03 \\
\bottomrule
\end{tabularx}
\end{table*}

\begin{table*}[htbp]
\centering
\footnotesize
\caption{CIR Taxonomy of Influencing Context Factors, with corresponding papers, and NIST CSF Elements from NIST SP 800-61r3.}
\label{tab:cirif_overview_context}
\begin{tabularx}{\textwidth}{p{4cm} L L}
\toprule
    \textbf{CIR-IF Taxonomy Category} &
    \textbf{Scientific Research} &
    \textbf{NIST SP 800-61r3} \\
\midrule\\[-3ex]
\categoryrow{Attacker}
\extendrow
        Motivation and Objectives &
        \cite{schlette_you_2024}, \cite{line_understanding_2015}, \cite{reed_simulation_2014}, \cite{pieterse_computer_2014} &
        none \\
\colorrow\extendrow
        Capabilities, TTP &
        \cite{kokulu_matched_2019}, \cite{backman_normal_2023}, \cite{line_understanding_2015}, \cite{reed_simulation_2014}, \cite{pieterse_computer_2014}, \cite{burkhead_phenomenological_2014} &
        none \\
\categoryrow{Regulations}
\extendrow
        Laws and official Regulations &
        \cite{bulgurcu_environmental_2024}, \cite{riebe_values_2024}, \cite{schlette_you_2024}, \cite{vilberth_security_2020}, \cite{pieterse_computer_2014}, \cite{akbas_enhancing_2023}, \cite{devey_triage_2019}, \cite{imrichova_gdpr_2020}, \cite{kamara_computer_2022}, \cite{mooi_prerequisites_2015}, \cite{pamnani_incident_2023} &
        GV.OC-03, RS.CO-02, RS.CO-03 \\
\colorrow\extendrow
        Industry Norms, Standards, Frameworks &
        \cite{stevens_how_2022}, \cite{schlette_you_2024}, \cite{line_understanding_2015}, \cite{vilberth_security_2020}, \cite{pieterse_computer_2014}, \cite{schlette_comparative_2021} &
        none \\
\categoryrow{Economic}
\extendrow
        Budget, Revenue &
        \cite{bulgurcu_environmental_2024}, \cite{sundaramurthy_tale_2014}, \cite{kokulu_matched_2019}, \cite{orellane_cybersecurity_2020}, \cite{vilberth_security_2020}, \cite{wiik_limits_2005}, \cite{cadena_metrics_2020}, \cite{mooi_management_2016}, \cite{mooi_prerequisites_2015}, \cite{wiik_chronic_2009} &
        none \\
\colorrow\extendrow
        Business Impact &
        \cite{colombo_escalation_2020}, \cite{mepham_impact-focused_2015}, \cite{shaked_operations-informed_2023} &
        GV.OC-01, GV.OC-02, GV.OC-04, GV.OC-05, GV.RM-01, GV.RM-03, GV.RM-04, ID.AM-05, ID.RA-03, ID.RA-04, ID.RA-05, DE.AE-04, RS.MA-03, RS.MA-04, RS.MA-05, RS.AN-08, RS.CO-02, RS.MI-01, RS.MI-02 \\
        
\extendrow
        Industry Partner, Supply Chain &
        \cite{bulgurcu_environmental_2024}, \cite{schlette_you_2024}, \cite{catota_cybersecurity_2018}, \cite{woods_how_2021}, \cite{burkhead_phenomenological_2014}, \cite{nyman_are_2019} &
        GV.RR-02, GV.SC-01, GV.SC-02, GV.SC-03, GV.SC-04, GV.SC-05, GV.SC-06, GV.SC-07, ID.AM-04, ID.RA-10, DE.CM-06, RS.CO-02, RS.CO-03, RS.MI-01, RS.CO-03 \\
\categoryrow{Technology}
\extendrow
        Tooling, Usability, Data, IT Infrastructure &
        \cite{nepal_burnout_2024}, \cite{bulgurcu_environmental_2024}, \cite{riebe_values_2024}, \cite{nyre-yu_identifying_2021}, \cite{kaufhold_cyber_2022}, \cite{sundaramurthy_turning_2016}, \cite{sundaramurthy_tale_2014}, \cite{kokulu_matched_2019}, \cite{schlette_you_2024}, \cite{al_sabbagh_cybersecurity_2019}, \cite{line_understanding_2015}, \cite{nyre-yu_observing_2019}, \cite{vilberth_security_2020}, \cite{wiik_limits_2005}, \cite{naseer_framework_2018}, \cite{pieterse_computer_2014}, \cite{ab_rahman_evidence-based_2016}, \cite{majid_model_2021}, \cite{akbas_enhancing_2023}, \cite{andrade_cognitive_2019}, \cite{cook_managing_2018}, \cite{graf_decision_2016}, \cite{happa_assessing_2021}, \cite{jaatun_framework_2009}, \cite{jaatun_empirical_2020}, \cite{pamnani_incident_2023}, \cite{smith_agile_2021}, \cite{wiik_chronic_2009} &
        ID.AM-*, PR.AA-01, PR.AA-02, PR.AA-03, PR.AA-04, PR.DS-*, PR.PS-*, PR.IR-*, DE.AE-02, DE.AE-03, DE.AE-06, DE.AE-07, RS.MA-01, RS.MI-02 \\
\categoryrow{Processes}
\extendrow
        Standard Operating Procedures, Playbooks, Policies&
        \cite{nepal_burnout_2024}, \cite{stevens_how_2022}, \cite{sundaramurthy_turning_2016}, \cite{sundaramurthy_tale_2014}, \cite{kokulu_matched_2019}, \cite{schlette_you_2024}, \cite{auzina_cyber_2023}, \cite{bitzer_managing_2023}, \cite{brown_incident_2016}, \cite{line_understanding_2015}, \cite{nyre-yu_observing_2019}, \cite{shaked_operations-informed_2023}, \cite{staves_cyber_2022}, \cite{tan_incident_2003}, \cite{vilberth_security_2020}, \cite{wiik_limits_2005}, \cite{naseer_framework_2018}, \cite{pieterse_computer_2014}, \cite{majid_model_2021}, \cite{burkhead_phenomenological_2014}, \cite{grispos_rethinking_2014}, \cite{irumudomon_qualitative_2024}, \cite{pamnani_incident_2023}, \cite{rollason-reese_incident_2003}, \cite{schlette_comparative_2021}, \cite{shaked_model-based_2022}, \cite{smith_agile_2021}  &
        GV.RM-06, GV.RR-02, GV.PO-01, GV.PO-02, GV.SC-08, GV.SC-09, GV.SC-10, ID.AM-*, ID.RA-06, ID.RA-09, ID.IM-04, PR.AA-05, PR.DS-*, PR.PS-06, DE.AE-02, DE.AE-08, RS.MA-01, RS.MA-02, RS.MA-03, RS.MA-05, RS.AN-03, RS.AN-06, RS.AN-07, RS.AN-08, RS.CO-02, RS.CO-03, RS.MI-01, RS.MI-02, RC.RP-* \\
\categoryrow{Organization}
\extendrow
        Structure, Size, Hierarchy &
        \cite{nepal_burnout_2024}, \cite{sundaramurthy_turning_2016}, \cite{sundaramurthy_tale_2014}, \cite{kokulu_matched_2019}, \cite{schlette_you_2024}, \cite{al_sabbagh_cybersecurity_2019}, \cite{auzina_cyber_2023}, \cite{line_understanding_2015}, \cite{nyre-yu_observing_2019}, \cite{vilberth_security_2020}, \cite{bartnes_line_current_2016}, \cite{grobler_common_2010}, \cite{jaatun_empirical_2020}, \cite{killcrece_organizational_2003}, \cite{killcrece_state_2003} &
        none \\
\colorrow\extendrow
        Management: People, HR, Daily Work, Work-Load &
        \cite{nepal_burnout_2024}, \cite{bulgurcu_environmental_2024}, \cite{sundaramurthy_human_2015}, \cite{stevens_how_2022}, \cite{sundaramurthy_turning_2016}, \cite{sundaramurthy_tale_2014}, \cite{kokulu_matched_2019}, \cite{webb_organizational_2017}, \cite{schlette_you_2024}, \cite{ahmad_how_2021}, \cite{al_sabbagh_cybersecurity_2019}, \cite{auzina_cyber_2023}, \cite{brown_incident_2016}, \cite{pfleeger_improving_2017}, \cite{ruefle_computer_2014}, \cite{shaked_operations-informed_2023}, \cite{steinke_improving_2015}, \cite{vilberth_security_2020}, \cite{wiik_limits_2005}, \cite{acarturk_continuous_2021}, \cite{agyepong_systematic_2023}, \cite{albluwi_framework_2017}, \cite{bada_improving_2014}, \cite{buchler_sociometrics_2018}, \cite{cadena_metrics_2020}, \cite{chamkar_soc_2023}, \cite{grobler_common_2010}, \cite{ioannou_cybersecurity_2019}, \cite{mooi_management_2016}, \cite{mooi_prerequisites_2015}, \cite{rollason-reese_incident_2003}, \cite{wiik_chronic_2009} &
        GV.RR-03, GV.RR-04, PR.IR-04, DE.AE-02 \\
\extendrow
        Organizational Learning &
        \cite{kokulu_matched_2019}, \cite{ahmad_case_2015}, \cite{al_sabbagh_cybersecurity_2019}, \cite{garmon_study_2019}, \cite{humphrey_identifying_2017}, \cite{shaked_operations-informed_2023}, \cite{ahmad_incident_2012}, \cite{cho_capturing_2020}, \cite{grispos_rethinking_2014}, \cite{ho_consciousness_2021}, \cite{pamnani_incident_2023}, \cite{shedden_informal_2011} &
        GV.PO-02, GV.OV-01, GV.OV-02, GV.OV-03, GV.SC-09, ID.RA-07, ID.RA-08, ID.IM-01, ID.IM-02, ID.IM-03, PR.AT-01, PR.AT-02, RS.AN-03, RS.RP-06 \\
\colorrow\extendrow
        Company Culture \& Business Model &
        \cite{nepal_burnout_2024}, \cite{bulgurcu_environmental_2024}, \cite{sundaramurthy_turning_2016}, \cite{sundaramurthy_tale_2014}, \cite{webb_organizational_2017}, \cite{al_sabbagh_cybersecurity_2019}, \cite{padayachee_coordinated_2020}, \cite{steinke_improving_2015}, \cite{ahmad_incident_2012}, \cite{grobler_common_2010}, \cite{ioannou_cybersecurity_2019}, \cite{jaatun_empirical_2020} &
        GV.RM-02, GV.RM-07, GV.RR-01, RS.CO-02, RS.CO-03, RC.CO-03, RC.CO-04 \\
\bottomrule
\end{tabularx}
\end{table*}

%% file: tables/tbl_taxonomy_comparison.tex
\begin{table*}[htb]
\centering
\caption{Taxonomy Characteristics Comparison}
\label{tab:taxonomy_comparison}
\footnotesize%
\begin{tabularx}{\textwidth}{@{}r|ccccccc@{}}
    \toprule
    \thead{\textbf{CIR-IF Taxonomy (ours)}} &
    \thead{\textbf{CSIRT MM}\cite{mooi_management_2016}} & 
    \thead{\textbf{PPTGC}\cite{vilberth_security_2020}} &
    \thead{\textbf{CIRF}\cite{pieterse_computer_2014}} &
    \thead{\textbf{Dynamic CIR}\cite{naseer_framework_2018}} &
    \thead{\textbf{AIR4ICS}\cite{smith_agile_2021}} &
    \thead{\textbf{IR Perf.}\cite{pfleeger_improving_2017}} &
    \thead{\textbf{IRM3}\cite{bitzer_managing_2023}} \\
    \midrule 
    Knowledge    & \harveyBallNone[0.75ex]
                 & \harveyBallNone[0.75ex]
                 & \harveyBallNone[0.75ex]
                 & \harveyBallHalf[0.75ex]
                 & \harveyBallHalf[0.75ex]
                 & \harveyBallFull[0.75ex]
                 & \harveyBallFull[0.75ex] \\
    CPTB         & \harveyBallNone[0.75ex]
                 & \harveyBallNone[0.75ex]
                 & \harveyBallNone[0.75ex]
                 & \harveyBallHalf[0.75ex]
                 & \harveyBallHalf[0.75ex]
                 & \harveyBallFull[0.75ex]
                 & \harveyBallHalf[0.75ex] \\
    \midrule
    \midrule
    Attacker     & \harveyBallNone[0.75ex]
                 & \harveyBallNone[0.75ex]
                 & \harveyBallHalf[0.75ex]
                 & \harveyBallFull[0.75ex]
                 & \harveyBallNone[0.75ex]
                 & \harveyBallNone[0.75ex]
                 & \harveyBallNone[0.75ex] \\
    Regulations  & \harveyBallFull[0.75ex]
                 & \harveyBallHalf[0.75ex]
                 & \harveyBallFull[0.75ex]
                 & \harveyBallNone[0.75ex]
                 & \harveyBallNone[0.75ex]
                 & \harveyBallNone[0.75ex]
                 & \harveyBallNone[0.75ex] \\
    Economic     & \harveyBallFull[0.75ex]
                 & \harveyBallNone[0.75ex]
                 & \harveyBallHalf[0.75ex]
                 & \harveyBallFull[0.75ex]
                 & \harveyBallNone[0.75ex]
                 & \harveyBallNone[0.75ex]
                 & \harveyBallNone[0.75ex] \\
    Technology   & \harveyBallFull[0.75ex]
                 & \harveyBallFull[0.75ex]
                 & \harveyBallHalf[0.75ex]
                 & \harveyBallFull[0.75ex]
                 & \harveyBallHalf[0.75ex]
                 & \harveyBallNone[0.75ex]
                 & \harveyBallFull[0.75ex] \\
    Process      & \harveyBallFull[0.75ex]
                 & \harveyBallFull[0.75ex]
                 & \harveyBallHalf[0.75ex]
                 & \harveyBallFull[0.75ex]
                 & \harveyBallFull[0.75ex]
                 & \harveyBallNone[0.75ex]
                 & \harveyBallFull[0.75ex] \\
    Organization & \harveyBallHalf[0.75ex]
                 & \harveyBallHalf[0.75ex]
                 & \harveyBallHalf[0.75ex]
                 & \harveyBallHalf[0.75ex]
                 & \harveyBallHalf[0.75ex]
                 & \harveyBallHalf[0.75ex]
                 & \harveyBallFull[0.75ex] \\
    \bottomrule
\end{tabularx}

\vspace{0.5em} %
\begin{minipage}{0.9\linewidth} %
    \footnotesize %
    The first column denotes the categories defined in our CIR-IF Taxonomy. The remaining cells indicate the extent to which existing academic frameworks cover the corresponding category;
    \harveyBallFull[0.75ex] = fully covered; \harveyBallHalf[0.75ex] = partially covered; \harveyBallNone[0.75ex] = not covered.
\end{minipage}
\end{table*}

%% file: src/06_discussion.tex
\section{Discussion}
\label{sec:discussion}

Our systematic literature review classified the influencing factors in 105 academic publications from over two decades (RQ 1); see \cref{tab:cirif_overview_human} and \cref{tab:cirif_overview_context}.
By adapting \cite{nickerson_method_2013}, we were able to define each factor type by assigning a subcategory, thereby ensuring completeness in the context of the final set of analyzed articles from \textbf{Phase III}.

\subsection{Summary and Interpretation of Key Results}
The \textit{CIR-IF Taxonomy} (RQ 2) is the first structured, science-domain-agnostic approach for organizing influencing factors in incident response.
An application of the taxonomy for a Mayring content analysis is described in this article, and a code book with text examples is provided in the appendix.

Reflecting on the findings of our analysis (RQ 3), one insight is particularly striking: the important role of human factors, organizational structure, and business requirements is not covered in the current incident response standard NIST SP 800-61r3.
\cref{tab:taxonomy_comparison} shows that existing academic frameworks address only particular subsets of categories and, consequently, are less comprehensive than the taxonomy proposed in this work.
For the systematic planning of future research, it is essential that researchers have a comprehensive understanding of the factors affecting the various dimensions of an organization’s incident response capability.

The National Institute of Standards and Technology (NIST) is responsible for standardization in physical science, engineering, and information technology.
Therefore, it is clear that the focus is on technology-related topics rather than on management and organizational requirements.

On the other hand, the senior security manager classified several NIST CSF elements as \textit{Business Impact}. However, the research team identified only a few factors in this category.
This could also indicate a lack of scientific coverage and represents a worthwhile field for future studies.
The high number of CSF elements assigned to the human-related factor \textit{Cybersecurity Situational Awareness} stems from the fact that technical data, such as asset management data (ID.AM) or vulnerability management data (IA.RA), were considered prerequisites for cognitive situational awareness.

\subsection{Takeaways from Literature Review}
Through the systematic review of the existing literature, we identified a set of factors associated with organizational communication processes and interpersonal collaboration.
While technology, processes, and organizational learning are comparatively well understood, human-centered factors such as leadership and team management should not be overlooked. In particular, ineffective communication, unclear responsibilities, and insufficient coordination can increase cognitive load and uncertainty, contributing to security-related stress. Prior work shows that such stress does not merely affect job satisfaction or efficiency but can fundamentally alter how individuals approach security tasks~\cite{darcyUnderstandingEmployeeResponses2014}.
In this section, we first provide a concise synthesis of the principal findings, followed by citations to the relevant studies included in our analysis.

\subsubsection{Trust and Empowerment}
Trust and empowerment appear to constitute fundamental human-related influence factors, mentioned in several of the examined empirical studies.
For example, specialists need the right privileges in advance to work to their full potential (see below).
To identify obstacles that need to be eliminated in daily work, it is important for managers to listen and thereby gain their employees' trust.
This also has the added benefit of making it easier to resolve conflicts as they arise and ensuring that employees' mental health does not suffer.
Another positive side effect might be lower staff turnover.

In 2024, a Microsoft research showed that empowerment, in addition to other factors, has an impact on mental health~\cite{nepal_burnout_2024}: "Empowerment is an individual’s sense of authority and autonomy for decision-making and accomplishing job expectations. When employees perceive that they have the capacity and authority to take actions to influence decisions that affect their work, it leads to better job engagement and lower levels of burnout [...]"

The study by Sundaramurthy et al.~\cite{sundaramurthy_human_2015} shows that security analysts are often insufficiently empowered, which negatively affects morale and effectiveness. Even highly skilled analysts may lack necessary privileges due to managerial concerns. For example, one senior analyst pointed out that his manager was reluctant to grant privileged access to his team for liability reasons. Empowerment was crucial for improving analysts’ morale, and SOC managers should explicitly integrate it into their organizational and managerial practices.

In another study, Sundaramurthy et al.~\cite{sundaramurthy_turning_2016} highlight that trust between managers and analysts is crucial for effective conflict resolution, as managers must actively support and enable operational changes.
"Above all, managers should earn the trust of their analysts and be a participant in the conflict resolution process as they are the authoritative persons to bring actual changes
to operations." Similarly, Schlette et al.~\cite{schlette_you_2024} show that insufficient trust and restrictive permission structures can hinder timely incident response. One participant noted that analysts sometimes need to act without explicit approval when a rapid response is required.

\subsubsection{Conflicts and Misplaced Priorities}
Another factor that can affect an organization's efficiency is misplaced priorities in performance measurement.
In SOC environments, analysts are often evaluated using KPIs that emphasize short-term output, such as the number of incidents resolved per day. While these metrics are easy to quantify, meeting these targets consumes much of an analyst's time and incentivizes them to maximize throughput, even when they recognize the need for systematic improvements (e.g., automation or root cause analysis). This misalignment introduces cognitive overload and goal conflict, as analysts must continuously trade off between immediate performance targets and long-term improvements. Under such conditions, individuals tend to adopt coping strategies that prioritize measurable outcomes over less visible but strategically important activities. Prior work shows that stress induced by complex, ambiguous, and burdensome security requirements leads individuals to adopt emotion-focused coping strategies, rather than problem-focused ones~\cite{darcyUnderstandingEmployeeResponses2014}. In such situations, employees are more likely to engage in moral disengagement (e.g., closing tickets to meet KPIs without conducting root cause analysis). Importantly, this shift is not driven by malicious intent, but by the need to cope with persistent stress and conflicting demands. This mechanism reinforces the status quo, leading to an accumulation of unresolved root causes and insufficient investment in automation, ultimately degrading the organization’s long-term security posture.

We recommend using a KPI that measures improvement within the SOC (e.g., a reduction in recurring incidents or an increase in automation coverage). This affects not only security teams but also creates conflicting priorities between departments. One department may achieve its goals, while the other may not. We argue that these conflicts must be made transparent and must be resolved so that all departments can achieve their goals and the entire organization can benefit.

The following analyzed studies make these problems clear: analysts have too many incidents to work on~\cite{nepal_burnout_2024,bulgurcu_environmental_2024}, and managers often measure their analysts' performance with inadequate metrics~\cite{sundaramurthy_turning_2016,kokulu_matched_2019} that cause conflicts, misaligned priorities, and stress.
An interview participant in \cite{kokulu_matched_2019} expressed it clearly: "'A useless metric can be the number of events because generally, people will not take into account false positive numbers [...] I think the reason is that the lack of understanding of security of the upper-class management as well as, it is a way to make it seem like your SOC is improving.'"

Webb et al. \cite{webb_organizational_2017} identified differing priorities and conflicting objectives among teams as a severe hindrance to \textit{Organization Security Learning}. "The NIRT and HIRT were primarily concerned with restoring IT services, with little or no time allotted for deeper incident analyses."

\subsubsection{Collaboration and Communication}
Collaboration and communication between and within teams are important to encourage information and knowledge sharing, build trust, and, consequently, improve cooperation.

Frequent and honest communication \cite{kokulu_matched_2019, schlette_you_2024} between shifts, teams, and organizational units enables collaboration and increases knowledge and awareness \cite{webb_organizational_2017}.
The user study from Kokulu et al.~\cite{kokulu_matched_2019} contains the following interview statement:" P18 mentioned the lack of information maintenance and the lack of communication between IT operations and security teams: '(The information the SOC requests) is not always up to date. [...] (The SOC’s receiving alerts from unknown devices) is usually because the IT ops and security teams are not necessarily the same people, and messages get lost and things like that.'"

Other statements indicate that collaboration among analysts, teams, and managers can overcome barriers \cite{sundaramurthy_turning_2016, bulgurcu_environmental_2024} and is crucial for enabling informed decision-making -- particularly since corporate culture can significantly hinder effective incident response.

\subsection{Potential cross-study Factor Relation Proposal}
The \textit{CIR-IF Taxonomy} is based on the reasons described in the statements or on the findings of 105 studies.
A reason, sometimes described in relation to a perceived result or as an empirical result of the corresponding study, builds the influencing factor.
Using the same categories for the results described for an influencing factor (see \cref{ssec:elementsofthetaxonomy}), we were able to construct possible chains linking the findings of different studies. 
\textbf{Note}: This kind of connection does not represent causality and was not empirically verified; the described chains can support future research efforts.

The construction of a chain begins by evaluating factors as described in \cref{sec:methodology:relationsship}.
The initial connection in the sequence links the selected factor to the subcategory.
To develop further connections, the analysis examines the influence of additional factors on various subcategories within this original subcategory.
For clarity and precision, a specific influencing factor and its directly affected subcategory are selected to create each link in a chain.
Multiple chains can be constructed by selecting different influencing factors for each subcategory.
The chain terminates as soon as our data indicate that there are no more relationships for a subcategory.
Our example illustrates a chain up to this self-contained point.

\input{img/chains}

\cref{fig:chains} shows a possible chain that starts with an influencing factor from the \textit{Management} subcategory.
The impact of the selected influencing factor is a reduction in funding due to the lack of effective metrics that communicate the SOC's value.
This has a direct impact on the SOC budget.
The \textit{Budget, Revenue} subcategory of the taxonomy lists the effects related to funding.
One of the influences in this subcategory is that a budget that is too small leads to poorly qualified analysts. 
According to an influencing factor in the \textit{Knowledge, Hard and Soft Skills} subcategory, inadequate analyst skills directly impact analyst burnout.
This leads to the \textit{Stress Handling} subcategory.
In this subcategory, according to our data, influencing factors affect only the same subcategory. 
Therefore, the chain terminates there.
This dependency chain shows an indirect impact of metrics that communicate the value of an SOC on personal well-being or analysts' burnout.
The influencing factor \textit{"Burnout among analysts, which is related to the lack of budget"} \cite{kokulu_matched_2019} supports this dependency chain. It acts as a shortcut between the \textit{Budget, Revenue} and the \textit{Stress Handling} subcategories.

The identified dependencies are consistent with the well-established Job Demands–Resources (JD-R) model~\cite{demerouti_job_2001}, which posits that burnout results from an imbalance between job demands and resources.
According to this framework, job resources, such as opportunities for training and development, autonomy, social support, feedback, and access to knowledge and material resources, are essential to mitigating the negative impact of job demands on employees' well-being.
When these resources are insufficient or absent, employees struggle to meet work requirements, leading to emotional exhaustion and disengagement. 
Thus, the indicated relationship between limited organizational resources and elevated levels of burnout in our possible chain is supported by the JD-R model’s prediction that a shortage of job resources is a key influence factor of burnout.

Mapping the influencing factors allowed us to connect factors across multiple articles, providing a tool for researchers and practitioners to identify possible dependencies that could serve as a starting point for generating theories or for understanding the dynamics within the organization.

\subsection{Limitations and Future Research}
The literature selection process encompassed an wide historical time span and, consequently, a large number of publications.  
The research design, driven by our research questions, and the quality-oriented analytical approach resulted in several limitations.

\textbf{Completeness.} The scope of the literature search and the reliance on English, peer-reviewed sources may have excluded relevant gray literature or non-English publications.
While we sought theoretical saturation, additional studies could introduce new factors influencing the construct or refine existing categories.
Our literature search was finished in May 2024; thus, more recent developments may not be fully captured here.
Missing categories can be added to the taxonomy following this methodology, as described in our work.
The taxonomy reflects the current state of academic research and may need to be adapted as incident response practices evolve.
Furthermore, the resulting taxonomy may primarily reflect academic perspectives rather than industry practices.
The selected scientific work we analyzed varies in research approaches, data types, sample sizes, and cohorts.
Articles with a strong research focus on a specific topic, i.e., human-computer interaction (HCI), will increase the number of influencing factors, which creates a strong accumulation of factors in this area.
In addition, the applied selection and deselection criteria shaped the set of articles analyzed in this research.
This must be taken into account when interpreting this study.

\textbf{Objectiveness.} Qualitative coding and taxonomy construction involve subjective interpretation, even though multiple researchers and a verification procedure were used to mitigate bias.
A good understanding of grounded research that identifies the influencing factors and prepares them by providing context and breaking them down to reduce ambiguity, as well as conducting cross-checks and calculating inter-rater reliability, helps but is not a guarantee.
Therefore, our research method described in this study strives for high quality and transparency.
Our objective was to reduce biases as much as possible by comparing the results of three researchers, fostering critical thinking within the team, and openly discussing different views.

\textbf{Comparison with NIST.} We have chosen the NIST SP 800-61r3 industry standard for comparison with the \textit{CIR-IF Taxonomy} because it is well developed, widely used, and received a major update at the time of the study.
Mapping the elements of NIST SP 800-61r3 to the \textit{CIR-IF Taxonomy} was performed by a cybersecurity expert with more than two decades of professional experience in cyber defense management.
For the purposes of our study, a cross-check was not planned, as our intention was to adequately depict the coverage and focus of an industry standard. %

\textbf{Causality.} Although potential influencing factors were identified, their specific impact could not be quantitatively assessed.
We have synthesized existing knowledge and established a theoretical foundation to guide future research. 
The findings presented in this study should not be confused with scientifically proven causality. 
These chains can only be hints towards generating new theories.
The goal is to identify potential dependencies between factors from the existing knowledge base, dependencies that show a possible path for follow-up research.

\textbf{Future research.} The human-factors identified in this work could lead to future research to develop and verify the \textit{Knowledge, Skills, Abilities} (KSA) needed by incident response specialists.
KSA capabilities can support human resources departments and managers in training and hiring the best-fit incident response specialist.
Another avenue to advance existing research is to examine the potential influence of industry partners (context factors), such as insurance agencies and lawyers~\cite{woods_how_2021}, on collaboration and decision-making (human factors).
This influence could extend to misplaced priorities and KPIs for incident response teams, potentially exacerbating occupational stress and interpersonal conflicts~\cite{nepal_burnout_2024,bulgurcu_environmental_2024,sundaramurthy_turning_2016,kokulu_matched_2019,webb_organizational_2017}.

%% file: img/chains.tex
\begin{figure*}[htb]
  \centering
  \includegraphics[width=\textwidth]{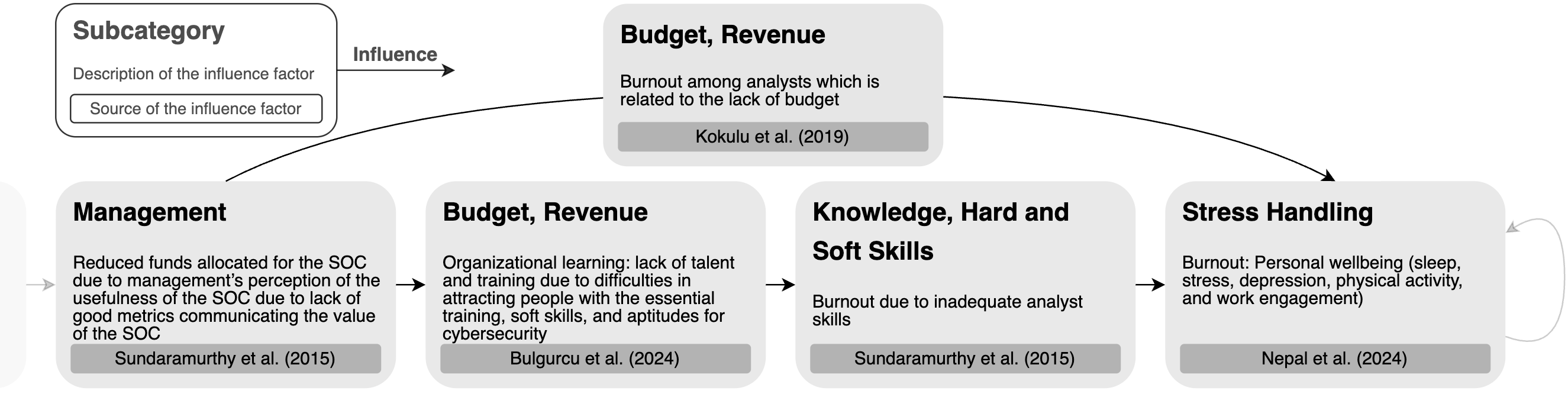} 
    \caption{Flowchart showing a indicated dependency chain of linked influencing factors. The chain begins in \textit{Management} and ends in \textit{Stress Handling}, which only influences itself, illustrated in both a direct and an extended version.}
  \label{fig:chains}
\end{figure*}

%% file: src/07_conclusions.tex
\section{Conclusion}
\label{sec:conclusion}
The body of research on cybersecurity incident response is growing rapidly, and the research community needs a clear, up-to-date framework to organize what we have learned.
We developed the first comprehensive, unified taxonomy of factors influencing cybersecurity incident response by systematically classifying and structuring the influencing factors identified in the research literature published over the past two decades.
In this study, we present the development of the \textit{CIR-IF Taxonomy}, which is designed to complement and extend NIST SP 800-61r3 and seven established academic frameworks.

For researchers, the \textit{CIR-IF Taxonomy} functions as a scalable conceptual map and analytical instrument for systematically characterizing and navigating the sociotechnical landscape of incident response within their empirical and theoretical investigations. 
Furthermore, we introduced a methodological approach to identifying categories of influencing factors and to supporting qualitative content analysis through a structured codebook.
The experimental linkage of influencing factors across multiple studies has the potential to facilitate theory building and to stimulate the generation of new research questions and hypotheses.
The consolidated and systematized body of knowledge presented in this work captures the current state of research on cybersecurity incident response at the human and organizational levels.

%% file: src/08_acks.tex
\section*{Acknowledgments}
\label{sec:acks}
We thank the anonymous reviewers for their great feedback on the paper and our shepherd for supporting us in preparing the current version.
This research work was supported by the National Research Center for Applied Cybersecurity ATHENE.
This research was also supported by the Graduate School for Applied Research in North Rhine-Westphalia (Graduate School NRW).
Jonas Kaspereit was supported by the research project ’Cyber Security Incident Response für KMUs (CySIRK)’ (13FH101KB1) of the German Federal Ministry of Education and Research (BMBF) and the nicos AG.
Marius Brockhoff was supported by the research project "North-Rhine Westphalian Experts in Research on Digitalization (NERD II)", sponsored by the state of North Rhine-Westphalia – NERD II 005-2201-0014.
Lea Gröber conducted a portion of this work while supported by a fellowship within the IFI program of DAAD, funded by the German Federal Ministry of Education and Research (BMBF).
The authors thank Katharina Krombholz and her team at the CISPA Helmholtz Center for Information Security for their hospitality and constructive discussions.

%% file: src/09_appendix.tex
\appendices
\label{appendix:main}
\section{Systematic Literature Review}
\label{appendix:literature_review}

\subsection{Search Terms and Databases}
\subsubsection{Google Scholar}
For Google Scholar, we always used the parameters \texttt{exclude patents, include citations}. See table \ref{tab:sq_google_scholar} on page \pageref{tab:sq_google_scholar}.

\subsubsection{JSTOR}
For JSTOR, we always used the parameters \texttt{Access type: Everything, Language: English}. See table \ref{tab:sq_jstor} on page \pageref{tab:sq_jstor}.

\subsubsection{SCOPUS}
For the SCOPUS database, we always used the parameters \texttt{Subject area: Computer Science; Engineering; Social Science; Decision Science; Business, Management, and Accounting}. See table \ref{tab:sq_scopus} on page \pageref{tab:sq_scopus}.

\section{Terms and Definitions}
\subsection{Incident Response in Organizations}
A cybersecurity incident is "... an event compromising the availability, authenticity, integrity, or confidentiality of stored, transmitted, or processed data or of the services offered by, or accessible via, network and information systems" Art. 6 (6) EU Directive 2022/2555 \cite{NIS2DirectiveEU2022_2555}
Cybersecurity Incident Response (CIR), or short incident response, encompasses procedures, actions, techniques, and tactics to prevent, identify, analyze, contain, and recover from a cybersecurity incident.
Depending on the size and structure of an organization, different types of internal organizational units, such as centralized Computer Emergency Response Teams or distributed Security Operations Centers, or external contractors, can be responsible for incident response.
The first Computer Emergency Response Team (CERT) was established in 1988 by an initiative of the Defense Advanced Research Projects Agency (DARPA) as a reaction to the \textit{Morris Worm} \cite{druffel_technical_2017}.
Cybersecurity incident response is a life-cycle process well described in NIST SP 800-61 rev. 3 \cite{nelsonIncidentResponseRecommendations2025a}.
CERTs and CSIRTs tend to have a reactive approach to incident handling, while Security Operations Centers (SOCs) can be seen as a proactive organizational unit. 

\subsection{Taxonomy}
A taxonomy is a systematic framework designed for the hierarchical classification of objects, concepts, or phenomena, based on their inherent relationships, shared characteristics, and defined boundaries.
It serves as a structured and common language that facilitates the clear organization, retrieval, and analysis of complex data within a particular domain.
The main objectives of a taxonomy are to ensure consistency and to provide a logical, mutually exclusive, and comprehensive categorization scheme.
Good examples are the Linnaean taxonomy for biological classification or the periodic table of the elements. 
Nickerson et al.~\cite{nickerson_method_2013} brought it to a point: \textit{"Much literature, however, uses taxonomy for systems of groupings that are derived conceptually or empirically."}

\subsection{Inter-Rater Reliability}
During taxonomy development and coding, Inter-Rater Reliability (IRR) is a common measure of quality.
Cohen's kappa ($\kappa$, \cite{cohen_coefficient_1960}) is a reliable statistic employed in social science and qualitative research to assess \textit{Inter-Rater Reliability} (IRR). This refers to the level of agreement between two independent raters (or coders) who classify items into distinct categories. Unlike a simple percentage agreement, Cohen's kappa is specifically designed to adjust for chance agreement.
\begin{math}
    \kappa = \frac{P_O - P_e}{1 - P_e}
\end{math}
; $P_O$ is the proportion of observed agreements, and $P_e$ is the proportion of expected agreements by chance.
The Landis \& Koch benchmark \cite{landis_measurement_1977} for interpreting Cohen's kappa value:
\input{tables/tbl_kappa_benchmark}

\section{Taxonomy Development}
\label{appendix:taxonomy_devel}

\subsection{Add an additional Layer: Adaption of formal Taxonomy Definition}
Formal description of a taxonomy, Nickerson et al. \cite{nickerson_method_2013}:
\begin{align*}
T = \big\{&D_i, i = 1, ..., n |\\ &D_i = \big\{C_{ij}, j = 1, ..., k_i; k_i \ge 2 \big\}\big\}
\end{align*}
In a taxonomy $T$, there are multiple dimensions $D_i$, where $i$ is the count, and each dimension has $k_i$ (at least two) characteristics (or categories in our case) $C_{ij}$.

During the development process, we introduced two levels of categories to improve usability.
Therefore, we adjusted the formal presentation of a taxonomy so that each category must have at least one subcategory $S$.

Adjusted formal description of a taxonomy with dimensions, categories, and subcategories:
\begin{align*}
T_{CIR-IF}= \big\{&D_i, i = 1, ..., p |\\ 
&D_i 
= \big\{C_{ij}, j = 1, ..., q_i |\\ 
&C_{ij} 
= \big\{S_{ijk}, k = 1, ..., r_{ij} \big\}\big\}\big\}
\end{align*}

\subsection{Taxonomy Example}
\cref{tab:CIRIF_taxonomy_example} is an example of the preliminary taxonomy and codebook.

\input{tables/tbl_CIR-IF-Taxonomy_example_01}

\section{Open Science}
To maximize the scientific and community value of our study, we provide a complete codebook with text examples, as well as an overview of the search queries used in our literature search, as part of this publication.
Furthermore, the BibTex files for phases I, II, and III, as well as the spreadsheets used during the taxonomy development process, are published at \url{https://anonymous.4open.science/r/research-IEEESPE}.

\section{Codebook}
The codebook is an ensemble of tables for human factors \cref{tab:if_codebook_human} and context factors \cref{tab:if_codebook_context_1}, \cref{tab:if_codebook_context_2}.
The first column lists the categories and subcategories of the \textit{CIR-IF Taxonomy}, which are used as labels in the coding process. The second column contains a compilation of influencing factors, which may include direct quotes from the research, abbreviated statements, or segmented statements, as detailed in \cref{ssec:elementsofthetaxonomy}.
These text examples work as a reference and can aid the coding process.
Occasionally, text examples start with a keyword like "Playbook:" or "Organizational Learning:". This initial word indicates the primary research topic from which the example is derived.

\input{tables/tbl_codebook}

\input{tables/tbl_litreview_phase1_google_scholar}
\input{tables/tbl_litreview_phase1_jstor}
\input{tables/tbl_litreview_phase1_scopus}

%% file: tables/tbl_kappa_benchmark.tex
\begin{table}[htbp]
\centering
\footnotesize
\label{tab:kappa_benchmark}
\begin{tabular}{@{}ccccc@{}}
\toprule
    Slight & Fair & Moderate & Substantial & almost Perfect \\
    .00 - .20 & .21 - .40 & .41 - .60 & .61 - .80 & > .80  \\ 
\bottomrule
\end{tabular}
\end{table}

%% file: tables/tbl_CIR-IF-Taxonomy_example_01.tex
\begin{table*}[htb]
\centering
\caption{Preliminary Taxonomy and Codebook Example}
\label{tab:CIRIF_taxonomy_example}
\small
\begin{tabularx}{\textwidth}{@{}lllX@{}}
\toprule
\thead{\textbf{Domain}} & \thead{\textbf{Category}} & \thead{\textbf{Subcategory}} & \thead{\textbf{Influencing Factor}} \\
\midrule
\midrule
Context & Organization & Management &
AT-Model Contradictions: This results in a number of metrices being defined to measure
the performance of SOC analysts. Ultimately, the job of the analysts is skewed very much
towards generating those defined metrices. This creates a conflict within them. They are confused
with two non-identical objectives – doing the right thing versus doing the required thing. \cite{sundaramurthy_turning_2016} \\
\midrule
Human & Knowledge & Employees' Knowledge &
Technology vs Needs: Expertise in situational context, \cite{nyre-yu_identifying_2021} \\
\bottomrule
\end{tabularx}
\end{table*}

%% file: tables/tbl_codebook.tex
\newcommand{\categoryrowshort}[1]{\rowcolor{gray!30}\multicolumn{2}{l}{\extendrow\textbf{#1}}\\[.1em]}

\begin{table*}[htbp]
\centering
\footnotesize
\caption{CIR-IF Taxonomy Codebook for Human Factors.}
\label{tab:if_codebook_human}
\begin{tabularx}{\linewidth}{p{4cm} L}
\toprule
    \textbf{Code Label / CIR-IF Category} &
    \textbf{Text Example / Influencing Factor} \\
\midrule\\[-3ex]
\categoryrowshort{Knowledge}
        Employees' Knowledge, Soft \& Hard Skills &
        Organizational learning: the lack of talent and training inhibits organizational learning \cite{bulgurcu_environmental_2024}; Technology vs Needs: Expertise in situational context \cite{nyre-yu_identifying_2021}; Complexity of the Threat Environment: limitations and difficulties in understanding, implementing, or using cybersecurity technologies within an organizational system that would protect against such advanced cyber attacks. \cite{bulgurcu_environmental_2024}; Technology: Nevertheless, our data unequivocally indicated that challenges pertaining to technology adaption are not primarily associated with the availability (or absence) of technology resources. Instead, these challenges are predominantly rooted in people and organization-specific factors, such as [a] insufficient talent and training, [...] \cite{bulgurcu_environmental_2024}; Burnout due to inadequate analyst skills \cite{sundaramurthy_human_2015}; Playbooks: Usability also depends on experience-level of the person using it \cite{stevens_how_2022}; Playbooks:  Quality also depends on [a] experience level, [...] \cite{stevens_how_2022}; Playbooks:  Quality also depends on [...][b] familiarity with playbook framework, [...] \cite{stevens_how_2022}; Playbooks:  Quality also depends on [...][d]  training and hands-on practice \cite{stevens_how_2022}; Inconsistent capability and maturity level among SOC teams \cite{kokulu_matched_2019}; Human Knowledge Issues: Insufficient analyst training \cite{kokulu_matched_2019})\\
        \colorrow\extendrow
Cybersecurity Situation Awareness &
        Organizational learning: employee naiveness or ignorance in regards to cybersecurity and their responsibilities, hindering an organization's learning capability due to the lack of cybersecurity awareness \cite{bulgurcu_environmental_2024}; Human Knowledge Issues: Low situational awareness \cite{kokulu_matched_2019}; Contextual knowledge and hunch feelings \cite{sundaramurthy_tale_2014})\\
\categoryrowshort{Cognition, Personality Traits, and Behavior}
\extendrow
        Intrinsic Motivation &
        Motivation of analysts \cite{sundaramurthy_tale_2014}; Lack of motivation among analysts \cite{kokulu_matched_2019})\\
\colorrow\extendrow
        Confidence &
        Over-confidence: three major areas employees' have excessive trust in, that are: a) cybersecurity professionals' perceived confidence in their knowledge and skill set, b) employees' perceived trust in technology [...] and c) employees' perceived trust in business processes - perceived confidence in existing operations proposed by the upper management \cite{bulgurcu_environmental_2024}; Stressors around incidents: Miscellaneous: Imposter syndrome \cite{nepal_burnout_2024} \\
\extendrow
        Stress Handling &
        Burnout: Personal well-being (sleep, stress, depression, physical activity, and work engagement) \cite{nepal_burnout_2024}; Burnout: Personal resources (Personality, Emotion Regulation, Resilience) \cite{nepal_burnout_2024}; Stressors around incidents: Miscellaneous: Lack of stress management skills \cite{nepal_burnout_2024}; Burnout due to lack of growth due to lack of creativity \cite{sundaramurthy_human_2015} )\\
\colorrow\extendrow
        Sensemaking &
        sensemaking addresses ‘what is going on?’... it can be translated into ‘is an incident going on?’ (detection) and if so, ‘what kind of incident is it?’(analysis). \cite{huis_human_2017}\\
\extendrow
        Decision-Making &
        It is an intuitive decision. It also depends per situation. For me, clear triggers that require me to reassess the situation are facts that indicate large data exfiltration, a dump of the active directory or indications that the attacker is changing his tactics. \cite{groenendaal_towards_2022}\\
\colorrow\extendrow
        Collaboration &
        AT-Model Contradictions: Lack empathy for other analysts: On the other hand, analysts have to constantly work with their colleagues in other roles, the lack of empathy creates barriers in this collaboration, thus fundamentally defeating the purpose of division of labor. \cite{sundaramurthy_turning_2016} \\
\bottomrule
\end{tabularx}
\end{table*}

\begin{table*}[htbp]
\centering
\footnotesize
\caption{CIR-IF Taxonomy Codebook for Context Factors (1/2).}
\label{tab:if_codebook_context_1}
\begin{tabularx}{\textwidth}{p{4cm} L}
\toprule
    \textbf{Code Label / CIR-IF Category} &
    \textbf{Text Example / Influencing Factor} \\
\midrule\\[-3ex]
\categoryrowshort{Attacker}
\extendrow
        Motivation and Objectives &
        Playbooks: Attacker behavior and motivation \cite{schlette_you_2024} \\
\colorrow\extendrow
        Capabilities, TTP &
        Inferior defense against specific types of attacks \cite{kokulu_matched_2019} \\
\categoryrowshort{Regulations}
\extendrow
        Laws and official Regulations &
        Playbooks: Laws and regulations \cite{schlette_you_2024}; Playbooks: Compliance by business structure – Business workflows necessitate compliance with regulatory requirements \cite{schlette_you_2024}; Playbooks: Privacy and data protection by location – Regional regulations, such as GDPR in the EU,  \cite{schlette_you_2024}; Playbooks: Security operations by sector – Industry-specific regulations \cite{schlette_you_2024}; Industry partners: Legal concerns: violation of laws due to collaboration as well as gray areas in legal frameworks \cite{bulgurcu_environmental_2024}; OSINT Technologies: IT infrastructures has to comply with regulations \cite{riebe_values_2024}; OSINT Technologies: privacy conflicts with training machine learning models \cite{riebe_values_2024}; OSINT Technologies: privacy regulations restricts potential efficiency gains \cite{riebe_values_2024} \\
\colorrow\extendrow
        Industry Norms, Standards, Frameworks &
        Playbooks: Industry standards and guidelines \cite{schlette_you_2024}; Playbooks: Usability and design (detail and completeness, visualization, implied tasks, correctness, precise language, delivering critical information without overwhelming the analyst, not enough detail/contingencies, not too many options for the analysts to follow, communication: missing list of essential contact points) also depends on features of the used framework (IACD, NIST) \cite{stevens_how_2022}; Playbooks: Perhaps the most significant drawback, observed in all phases of our evaluation, is that the frameworks do not elicit playbooks written in sufficient detail for real-world use. \cite{stevens_how_2022})\\
\categoryrowshort{Economic}
\extendrow
        Budget, Revenue &
        Organizational learning: lack of talent and training due to difficulties in affording people with the right talent \cite{bulgurcu_environmental_2024}; The challenge of constant justification of the SOCs value due to lack of profit margins \cite{sundaramurthy_tale_2014}; Operational Issues: Insufficient budget given by management \cite{kokulu_matched_2019}; Burnout among analysts which is related to the lack of budget \cite{kokulu_matched_2019} \\
\colorrow\extendrow
        Business Impact &
        Despite the acknowledged need for reflecting the impact of IR on operations, IR is typically executed only with technical considerations in mind, failing to consider the associated impact on business operations \cite{shaked_operations-informed_2023}; what factors become of influence when considering the transition between cyber incident management and cyber crisis management in a corporate environment \cite{colombo_escalation_2020};Decision-making is directly influenced by overall impact on the targeted organisation \cite{mepham_impact-focused_2015} \\
        
\extendrow
        Industry Partner, Supply Chain &
        Playbooks: Collaboration with business partners, when necessary \cite{schlette_you_2024}; Industry partners: companies with resources having the minimal incentive to help others as they also fear losing their competitive advantage \cite{bulgurcu_environmental_2024}; Industry partners: [...] our research findings revealed the existence of numerous gray areas and lack of incentives that hinder organizations from engaging in sharing and collaboration with one another. \cite{bulgurcu_environmental_2024}; Industry partner: One of the reasons for the lack of sharing has been identified as the vague nature of the relationship among parties that are business partners. \cite{bulgurcu_environmental_2024})\\
\categoryrowshort{Technology}
\extendrow
        Tooling, Usability, Data, IT Infrastructure &
        Playbooks: Centralization and Diversity of IT Infrastructure \cite{schlette_you_2024}; Stressors around incidents: Miscellaneous: Tools and processes that do not work as expected \cite{nepal_burnout_2024}; Playbooks: Resource availability by log data storage constraints – Incident handling adapts to storage limitations for log data, data retention \cite{schlette_you_2024}; Playbooks: Security tools \cite{schlette_you_2024}; Playbooks: Automation relies heavily on tools \cite{schlette_you_2024}; Technological Issues: poor usability of SOC systems reduce efficiency, when tools are difficult to master or lack integration it reduces efficiency \cite{kokulu_matched_2019}; Technological Issues: Malfunctioning of SOC tools \cite{kokulu_matched_2019}; Technological Issues: Insufficient automation Level of SOC Components \cite{kokulu_matched_2019}; Technological Issues: High false positive rate does NOT have a major impact on SOC operations \cite{kokulu_matched_2019}; Technological Issues: Poor quality of reports and logs \cite{kokulu_matched_2019}; Operational Issues: low visibility in network infrastructure \cite{kokulu_matched_2019}; Operational Issues: low visibility on endpoints \cite{kokulu_matched_2019}; Technology: issues related to dependencies on specific tools, over dependence on one product \cite{bulgurcu_environmental_2024}; OSINT Technologies: automation improves efficacy \cite{riebe_values_2024}; OSINT Technologies: accuracy creates trust, inaccurate systems are not trusted \cite{riebe_values_2024}; Technological Issues: Overloaded and low-quality Threat Intelligence \cite{kokulu_matched_2019}; OSINT Technologies: transparency (e.g., ML Algorithms) creates trust \cite{riebe_values_2024}; Technology vs Needs: Communication and collaboration \cite{nyre-yu_identifying_2021}; Dashboard design: customizability is important for good dashboards \cite{kaufhold_cyber_2022}; Dashboard design: tools need to operate on real-time data \cite{kaufhold_cyber_2022}; Dashboard design: cross-references (e.g., CVE linked across feeds) enables better situational awareness \cite{kaufhold_cyber_2022}; Dashboard design: Interoperability of tools improves workflow efficiency \cite{kaufhold_cyber_2022}; Dashboard design: Exporting reports supports communication \cite{kaufhold_cyber_2022}; AT-Model Contradictions: The bad integration of the tools leads to missing information about an alarm or incident leading to inefficient decision-making of the analyst ("poor attribution") \cite{sundaramurthy_turning_2016}; AT-Model Contradictions: In order to achieve the goal of division of labor, where analysts perform the tasks they are good at, it is imperative that they have the right tools to assists them. \cite{sundaramurthy_turning_2016}; Use of security tools and software \cite{sundaramurthy_tale_2014}; The workflow is directly influenced by the type of tools \cite{sundaramurthy_tale_2014})
\\
\categoryrowshort{Processes}
\extendrow
        Standard Operating Procedures, Playbooks, Policies &
        Playbooks: Process improvement by playbook development and automation \cite{schlette_you_2024}; Playbooks: Usability of playbooks by playbook customization and maintenance \cite{schlette_you_2024}; Playbooks: Workflow precision by internal IR directives \cite{schlette_you_2024}; Playbooks: Can help make up for lack of experience \cite{stevens_how_2022}; AT-Model Contradictions: The purpose of SOPs is to make sure for a given incident every analyst will respond in a similar way. In other words, they ensure predictability in operations. However, there is a fundamental conflict that SOPs face, which is between expected behavior and creativity of the analysts. While SOPs can empower an analyst within limits, the same SOPs can dis-empower the analyst from acting beyond them. \cite{sundaramurthy_turning_2016}; Operational workflows \cite{sundaramurthy_tale_2014}; Stressors around incidents: Documentation \cite{nepal_burnout_2024}; ineffective legal and corporate communications in case of an incident \cite{kokulu_matched_2019} \\
\bottomrule
\end{tabularx}
\end{table*}

\begin{table*}[htbp]
\centering
\footnotesize
\caption{CIR-IF Taxonomy Codebook for Context Factors (2/2).}
\label{tab:if_codebook_context_2}
\begin{tabularx}{\textwidth}{p{4cm} L}
\toprule
    \textbf{Code Label / CIR-IF Category} &
    \textbf{Text Example / Influencing Factor} \\
\midrule\\[-3ex]
\categoryrowshort{Organization}
\extendrow
        Structure, Size, Hierarchy &
        Playbooks: Security team-related factors: chain of command \cite{schlette_you_2024}; Playbooks: Security team-related factors: team size \cite{schlette_you_2024}; Playbooks: Security team-related factors: responsibilities \cite{schlette_you_2024}; AT-Model Contradictions: There is a dualism within division of work that leads to efficiency problems. The very specific role assignments to analysts leads to analysts working in silos, thus they often lack empathy for other analysts. \cite{sundaramurthy_turning_2016}; Team structure and hierarchy \cite{sundaramurthy_tale_2014}; Stressors around incidents: Miscellaneous: Working remotely \cite{nepal_burnout_2024}; Operational Issues: slow responding customers of outsourced SOC has an impact on the SOC's response speed \cite{kokulu_matched_2019}; Industry Susceptibility (e.g. airlines are easy targets for phishing (check-ins etc.)) \cite{kokulu_matched_2019})\\
\colorrow\extendrow
        Management: People, HR, Daily Work, Work-Load &
        Organizational learning depends on level of criticality of the incident (only critical incidents force learning, rational: only management can decide to learn from lesser critical incidents and make official in a IR policy) \cite{webb_organizational_2017}; Technology: Nevertheless, our data unequivocally indicated that challenges pertaining to technology adaption are not primarily associated with the availability (or absence) of technology resources. Instead, these challenges are predominantly rooted in people and organization-specific factors, [...][b] inadequate allocation of resources, [...] \cite{bulgurcu_environmental_2024}; Technology: Nevertheless, our data unequivocally indicated that challenges pertaining to technology adaption are not primarily associated with the availability (or absence) of technology resources. Instead, these challenges are predominantly rooted in people and organization-specific factors, such as [..][c] executive support, [...] \cite{bulgurcu_environmental_2024}; Technology: Nevertheless, our data unequivocally indicated that challenges pertaining to technology adaption are not primarily associated with the availability (or absence) of technology resources. Instead, these challenges are predominantly rooted in people and organization-specific factors, such as [..][d] or suboptimal decision-making processes. \cite{bulgurcu_environmental_2024}; Playbooks: Preparation needed (time to think, proactive planning, identifying the initiating condition) \cite{stevens_how_2022}; AT-Model Contradictions: The analysts were stuck performing a high volume repetitive task. Neither the analysts nor the field workers could invest time in any creative security projects [...] \cite{sundaramurthy_turning_2016}; Shift scheduling \cite{sundaramurthy_tale_2014}; Burnout: Job demands lead to stress and exhaustion \cite{nepal_burnout_2024}; Burnout: Job content (job security and skill discretion, empowerment and happiness at work, workload management, work-life balance, and boundaries) \cite{nepal_burnout_2024}; Stressors around incidents: Workload \cite{nepal_burnout_2024}; Stressors around incidents: Timing \cite{nepal_burnout_2024}; Stressors around incidents: Training \cite{nepal_burnout_2024}; Stressors around incidents: Priorities \cite{nepal_burnout_2024}; Stressors around incidents: Miscellaneous: Repetition of task \cite{nepal_burnout_2024}; motivation of analysts by career progression \cite{sundaramurthy_tale_2014}; Organizational learning: lack of talent and training due to difficulties in attracting people with the essential training, soft skills, and aptitudes for cybersecurity \cite{bulgurcu_environmental_2024}; Stressors around incidents: Miscellaneous: No recognition \cite{nepal_burnout_2024}; Stressors around incidents: Support \cite{nepal_burnout_2024}; Burnout: Social resources (Teamwork and team support, Role clarity, Recognition) \cite{nepal_burnout_2024}; Burnout due to lack of empowerment by management \cite{sundaramurthy_human_2015}; Reduced funds allocated for the SOC due to management's perception of the usefulness of the SOC due to lack of good metrics communicating the value of the SOC \cite{sundaramurthy_human_2015}; Burnout due to low creativity levels by lack of empowerment by management \cite{sundaramurthy_human_2015}; Human capital (skills, growth, empowerment, creativity) is positively influenced by an efficient SOC \cite{sundaramurthy_human_2015}; SOC efficiency is directly influenced by human capital(skills, growth, empowerment, creativity) \cite{sundaramurthy_human_2015}; AT-Model Contradictions: This results in a number of metrics being defined to measure the performance of SOC analysts. Ultimately, the job of the analysts is skewed very much towards generating those defined metrics. This creates a conflict within them. They are confused with two non-identical objectives - doing the right thing versus doing the required thing. \cite{sundaramurthy_turning_2016}; AT-Model Contradictions: There is also a primary contradiction within the objective of the SOC itself. Optimizing the work by reducing the number of security alerts would lead to a reduction of SOC analysts by the management. \cite{sundaramurthy_turning_2016}; AT-Model Contradictions: Managers must earn trust of their analysts to gain knowledge about contradictions \cite{sundaramurthy_turning_2016}; Inadequate metrics for measuring SOC efficiency \cite{sundaramurthy_tale_2014}; Lack of communication between managers and analysts \cite{kokulu_matched_2019}; Operational Issues: Inefficient evaluation metrics \cite{kokulu_matched_2019}; metrics are controversial between managers and analysts which causes disagreements and conflicts \cite{kokulu_matched_2019}; Playbooks: Quality also depends on [...][c] mentorship, [...] \cite{stevens_how_2022}; Complexity of the Threat Environment: Another severe consequence of this complexity (footnote: high frequency of incidents and vulnerabilities) is the decline in cyber professionals’ optimism, morale, and mental health. \cite{bulgurcu_environmental_2024}; Complexity of the Threat Environment: frequency of cyber incidents and their impact on the industry, leads to desensitization \cite{bulgurcu_environmental_2024}; Process and culture: analyst in conflict because of getting fired for finding an intruder \cite{bulgurcu_environmental_2024}; Playbooks: Security culture is influenced by communication by management support \cite{schlette_you_2024}; Playbooks: Security culture is influenced by permissions by management support, Permission and pre-authorization is important in case of an incident, act fast \cite{schlette_you_2024} \\
\extendrow
        Organizational Learning &
        phishing training of employees does not solve this most common attack \cite{kokulu_matched_2019} \\
\colorrow\extendrow
        Company Culture \& Business Model &
        Industry partners: lack of leadership support and not acknowledging the benefits of sharing \cite{bulgurcu_environmental_2024}; Communication between different day shifts \cite{sundaramurthy_tale_2014}; Process and culture: [...] another notable challenge resides in the prioritization of business continuity over enhancement of cybersecurity measures, relegation of security to an afterthought. \cite{bulgurcu_environmental_2024}; AT-Model Contradictions: [...] the commercial logic for having the tool is compliance not operational efficiency, resulting in this primary contradiction. \cite{sundaramurthy_turning_2016}; Barriers for organizational learning: Concerns about potential negative consequences of experimenting with new measures, operation inertia (tendency to rely on established practices) \cite{webb_organizational_2017}; Organizational learning: [...] the later admitted that such consultations [footnote: between management and response team during incident] would have benefited enterprise security objectives. \cite{webb_organizational_2017}; Organizational learning: hindered by different priorities and objectives between operation IR team and Information security management \cite{webb_organizational_2017}; Organizational learning: Integrating stakeholders is a prerequisite for org. learning \cite{webb_organizational_2017}; Organizational learning: due to communication gaps, the IS management has a lack of awareness \cite{webb_organizational_2017}; Stressors around incidents: Communication \cite{nepal_burnout_2024}; Industry partners: A predominant theme emerging from our research is the detrimental impact of the prevailing information sharing and collaboration culture within the industry. This culture is characterized by minimal practices of sharing and collaboration, which significantly hinder an organization's capacity to learn from cyber incidents. \cite{bulgurcu_environmental_2024} \\
\bottomrule
\end{tabularx}
\end{table*}

%% file: tables/tbl_litreview_phase1_google_scholar.tex
\begin{table*}[t]
    \rowcolors{1}{white}{gray!10}
    \renewcommand{\thead}[1]{\textbf{#1}}
    \footnotesize
    \centering
    \caption{Google Scholar: Search queries}
    \label{tab:sq_google_scholar}
    \begin{tabularx}{.95\linewidth}{l l L r}
        \toprule
         \thead{ID} & \thead{Date} & \thead{Query} & \thead{Results} \\
    \midrule\\[-3ex]
\extendrow
        ID\_20240504\_001 &
        2024-05-04 &
        security "incident response" influenc* factor &
        13,200 \\
\extendrow
        ID\_20240506\_001 &
        2024-05-06	&
        (security OR cyber) "incident response" "human?factor" influenc* &
        1,800 \\
\extendrow
        ID\_20240510\_001 &
        2024-05-10 &
        (security OR cyber) AND ("incident response" OR "incident handling" \linebreak  OR "incident management") AND influenc* AND "human?factor” &
        2,730\\
\extendrow
        ID\_20240512\_001 &
        2024-05-12 &
        (security OR cyber) AND ("eradication" OR "incident response" OR "incident handling" \linebreak  OR "incident management" OR "incident coordination") AND influenc* \linebreak AND "situation awareness” &
        2,420 \\
\extendrow
        ID\_20240515\_001 &
        2024-05-15 &
        (security OR cyber) AND ("eradication" OR "incident response" OR "incident handling" \linebreak OR "incident management" OR "incident coordination") AND influenc* \linebreak AND "decision making” &
        18,100 \\
\extendrow
        ID\_20240515\_002 &
        2024-05-15 &
        (security OR cyber) AND ("eradication" OR "incident response" OR "incident handling" \linebreak OR "incident management" OR "incident coordination") AND influenc* \linebreak  AND "threat intelligence” &
        3,910 \\
\extendrow
        ID\_20240516\_001 &
        2024-05-16 &
        (security OR cyber) AND ("eradication" \linebreak OR "incident response" OR "incident handling" \linebreak OR "incident management" \linebreak OR "incident coordination") AND influenc* AND "information sharing” &
        13,900 \\
\extendrow
        ID\_20240518\_001 &
        2024-05-18 &
        "computer emergency response" AND "situation awareness” &
        4,880 \\
\extendrow
        ID\_20240518\_001 &
        2024-05-18 &
        "computer emergency response" AND "influence factor" &
        8 \\
\extendrow
        ID\_20240518\_003 &
        2024-05-18 &
        "computer emergency response" AND "influencing factor” &
        33 \\
\extendrow
        ID\_20240518\_004 &
        2024-05-18 &
        (security OR cyber) AND ("eradication" OR "incident response" OR "incident handling" \linebreak OR "incident management" OR "incident coordination") AND influenc* \linebreak AND "motivation theory” &
        1,260 \\
\extendrow
        ID\_20240520\_001 &
        2024-05-20 &
        (security OR cyber) AND ("eradication" OR "incident response" OR "incident handling" \linebreak OR "incident management" OR "incident coordination") AND influenc* \linebreak AND "protection motivation theory” &
        804 \\
\extendrow
        ID\_20240520\_002 &
        2024-05-20 &
        "Computer Emergency Response Team" AND influenc* \linebreak AND "protection motivation theory” &
        72 \\
\extendrow
        ID\_20240520\_003 &
        2024-05-20 &
        (security OR cyber) AND ("eradication" OR "incident response" OR "incident handling" \linebreak OR "incident management" OR "incident coordination") AND influenc* \linebreak AND "context factor” &
        70 \\
\extendrow
        ID\_20240520\_004 &
        2024-05-20 &
        (security OR cyber) AND ("eradication" OR "incident response" OR "incident handling" \linebreak OR "incident management" OR "incident coordination") AND influenc* \linebreak AND "artificial intelligence” &
        10,800 \\
\extendrow
        ID\_20240520\_005 &
        2024-05-20 &
        (security OR cyber) AND ("eradication" OR "incident response" OR "incident handling" \linebreak OR "incident management" OR "incident coordination") AND influenc* \linebreak AND "decision support” &
        8,800  \\
\extendrow
        ID\_20240522\_001 &
        2024-05-22 &
        (security OR cyber) AND ("eradication" OR "incident response" OR "incident handling" \linebreak OR "incident management" OR "incident coordination") AND influenc* \linebreak AND "fatigue” &
        8,910 \\
\extendrow
        ID\_20240523\_001 &
        2024-05-23 &
        (security OR cyber) AND ("eradication" OR "incident response" OR "incident handling" \linebreak OR "incident management" OR "incident coordination") AND influenc* \linebreak AND "dynamic capabilities” &
        963 \\
\extendrow
        ID\_20240525\_001 &
        2024-05-25 &
        (security OR cyber) AND ("eradication" OR "incident response" OR "incident handling" \linebreak OR "incident management" OR "incident coordination") AND influenc* \linebreak AND "psychological effect” &
        1,620 \\
\extendrow
        ID\_20240525\_002 &
        2024-05-25 &
        (security OR cyber) AND ("eradication" OR "incident response" OR "incident handling" \linebreak OR "incident management" OR "incident coordination") AND influenc* \linebreak AND "organizational theory” &
        1,660 \\
 \bottomrule
    \end{tabularx}
\end{table*}

%% file: tables/tbl_litreview_phase1_jstor.tex
\begin{table*}[ht]
    \renewcommand{\thead}[1]{\textbf{#1}}
    \footnotesize
    \centering
    \caption{JSTOR: Search queries}
    \label{tab:sq_jstor}
    \begin{tabularx}{.9\linewidth}{l l L r}
        \toprule
         \thead{ID} & \thead{Date} & \thead{Query} & \thead{Results} \\
        \midrule
        ID\_20240527\_001 &
        2024-05-27 &
        cty:journal (security OR cyber) AND ("incident response" OR "incident handling" \linebreak OR "incident management" OR "incident coordination") AND influenc* \linebreak NOT (traffic OR military OR Food OR medicine) &
        98 \\ 
    \bottomrule
    \end{tabularx}
\end{table*}

%% file: tables/tbl_litreview_phase1_scopus.tex
\begin{table*}[ht]
    \renewcommand{\thead}[1]{\textbf{#1}}
    \footnotesize
    \centering
    \caption{SCOPUS: Search queries}
    \label{tab:sq_scopus}
    \begin{tabularx}{.85\linewidth}{l l L r}
        \toprule
         \thead{ID} & \thead{Date} & \thead{Query} & \thead{Results} \\
        \midrule
        ID\_20240527\_001 &
        2024-05-27 &
        ( security OR cyber ) AND ( "incident response" OR "incident handling" \linebreak OR "incident management" OR "incident coordination" ) AND influence &
        41 \\ 
        \hline
    \end{tabularx}
\end{table*}